\def\Msun{M_{\odot}}

\def\etal{{\it et~al.\ }}

\def\ie{{\it i.e.,~}}

\def\ltsima{$\; \buildrel < \over \sim \;$}
\def\simlt{\lower.5ex\hbox{\ltsima}}
\def\gtsima{$\; \buildrel > \over \sim \;$}
\def\simgt{\lower.5ex\hbox{\gtsima}}

\documentstyle[aasms4,psfig]{article}
\tighten
\singlespace
 
\lefthead{Ferrara \& Tolstoy}
\righthead{Evolution of Dwarf Galaxies}
 
\begin{document}
  
\title{The Role of Stellar Feedback and Dark Matter\\
in the Evolution of Dwarf Galaxies}
	
\author{Andrea Ferrara$^1$ \& Eline Tolstoy$^2$}
\affil{
$^1$Osservatorio Astrofisico di Arcetri \\ 50125 Firenze, Italy
\\ E--mail: ferrara@arcetri.astro.it\\
$^2$European Southern Observatory   
\\D-85748 Garching bei M\"{u}nchen, Germany
\\ E--mail: etolstoy@eso.org\\}
\begin{abstract}
Supernova and multiple supernova events regulate several
structural properties of dwarf galaxies. In particular, they govern
the metal enrichment and the energy budget of the ISM;
they might induce partial (blowout) or total (blowaway) gas removal
from the galaxy; they also regulate the pressure of the ISM and consequently
the morphology of the galactic gaseous body.
Significant amounts of dark matter may play an equally
important role: the dark matter gravitational potential
tends to concentrate baryons towards the center, thus enhancing
both the star formation rate and metal production. Also, the
dynamical properties of the ISM, and the occurrence 
of a blowout or blowaway are shown to be determined by the 
dark matter content.
We present detailed analytical/numerical models describing the evolution
of dwarf Irregular galaxies including the above and other effects. 
The main results are: 
{\it (i)} dwarfs with total masses $M\simlt 5\times 10^6~M_\odot$ are blown 
away;
those with gas masses up to $\simeq 10^9~M_\odot$ lose mass in an outflow;
{\it (ii) } metallicities are found to correlate tightly with dark matter 
content and 
are consistent with a range of  dark-to-visible mass ratios $\phi\approx 0-30 
$ 
with about 65\% of the dwarfs in the sample having $\phi\approx 0-10$; 
{\it (iii)} we predict a lower limit to the oxygen abundance in dIs of 
$12+log(O/H)\approx 7.2$;
{\it (iv)} outflows are not particularly important for the metallicity evolution 
of dwarf galaxies and certainly less than star formation for
gas consumption; however, dwarfs with gas masses few $\times 10^8~M_\odot$ 
are shown to be the major pollutants of the IGM; 
{\it (v)} the ISM HI velocity dispersion correlates with metallicity and, 
indepentently
of dark matter, scales as $Z^{3.5}$. 
Specific comparisons with well studied dI galaxies, as for example Leo A,
yield excellent agreement with the data. Based on our results, we discuss a 
scenario in which late type and early type dwarfs had common progenitors in 
the past, 
but differences in their total mass forced these objects to follow different 
evolutionary paths. Therefore, we consider dI $\rightarrow$ dE transitions
occurring at present cosmic times as very unlikely.

\end{abstract}
  
\section{Introduction}

Nearby dwarf galaxies are classified in four basic types:  {\it (i)}
Dwarf Irregular (dI) galaxies are the most common type by number, and
are usually unstructured gas rich systems with varying levels of 
star formation occurring in a haphazard manner across the galaxy.
The velocity field of the HI gas in these systems is dominated by
random motions rather than rotation (Binney \& de Vaucouleurs 1981;
Staveley-Smith, Davies \& Kinman 1992; Lo, Sargent \& Young 1993),
with peak column densities (as high as 4 $\times~10^{21} \rm cm^{-2}$
or 12~$M_\odot$~pc$^{-2}$) similar to those of larger galaxies;  {\it
(ii)} Blue Compact Dwarf (BCD) galaxies are gas rich systems dominated
by very active star formation, and resembling massive HII
regions seen in large galaxies.  They are thought to be forming stars
at a rate which they can only maintain for a short period (Papaderos
\etal 1994); {\it (iii)} Dwarf Spheroidal (dSph) galaxies usually have
no gas in their center down to very low limits (Knapp \etal 1978),
although it has been suggested that the gas supply may reside outside
the central regions, driven there by supernova explosions and stellar
outflows (Puche \etal 1992, Carignan 1998; Bowen \etal 1997). 
The stellar distribution
of dSph is similar to that of globular clusters, although less 
centrally concentrated, but a detailed study of
the stellar population often reveals that several distinct bursts of
star formation have occurred in the past (e.g. Smecker-Hane \etal
1994). Finally, {\it (iv)} Dwarf Elliptical (dE) galaxies look 
similar to their namesake elliptical galaxies. It was previously
thought that they contain only old stars, but new data 
suggest a sequence of
several bursts of star formation, some of which may be quite recent,
are needed to explain the characteristics of their stellar population
(Ferguson \& Binngeli 1994, Babul \& Ferguson 1996, Han \etal 1997).  

Thus, broadly speaking, there are two classes of dwarf galaxies.
Late-type (dI and BCD) are star forming systems; the main difference
between dI and BCD galaxies seems to be found in their star formation 
rate - and there is strong evidence that these are
otherwise very similar systems (e.g Tolstoy 1998a; Taylor 1997).
The other class consists of early-type
systems (dSph and dE) which are typically not {\it currently} forming
stars. 
The difference between dSphs and dEs is not well defined, and lies 
only in the total mass of the system (dEs are typically more massive than
dSphs). 

Connecting the above classes of dwarfs in a unified evolutionary scenario
has been the subject of considerable study.
In fact, in spite of the qualitative differences among the
various types, a structural kinship of early and late
type dwarfs becomes evident at a more quantitative level.
Binggeli (1994) compared the two classes in
terms of their central surface brightness and disk scale length
vs. absolute blue magnitude for a Virgo cluster sample.  He found
that early and late types show essentially the same distribution in
such a parameter space.  Even before these results, the concept of a
transition of some kind from late to early types has been the
rationale for a number of studies suggesting different evolutionary
schemes. Einasto \etal (1974), and later on
Lin \& Faber (1983), pointed out that since early type dwarfs are
often found as satellites of large galaxies, ram pressure stripping in
such a high density environment could provide an efficient mechanism
to remove all the gas from a galaxy, although
Binggeli (1994) gives a list of arguments against it. 
The importance of the environment for
the evolution of galaxies has already been
emphasized by Lacey \& Silk~(1991). 
The major difficulty is to explain
the much lower surface brightness of dIs with respect to dEs. 
One scenario for the dI $\rightarrow$ dE transition
was proposed by Davies \& Phillips (1988), who postulated that the gas,
instead of being swept away by some transport mechanism, could 
have been consumed by astration.  As it will become
clear from our results, astration (as well as outflows) do not appear 
to be able to rid a galaxy of its ISM. 
An alternate hypothesis is that
dIs and dEs might have had common progenitors, but have evolved
differently. What are the reasons for such a behavior ?
At least two possibilities need to be considered: {\it (i)} a
difference in the dark matter (DM) content; {\it (ii)} the effects of the
environment.
As we will show in this paper the DM fraction must
strongly affects the evolution of a low mass galaxy and its star
formation history.

Given that the major differences between the two classes
reside in their star formation histories, 
it is important to determine
the effects of star formation on the
ISM. 
Among these effects, the most important one is very likely 
the energy injection from massive stars. 
Since dwarfs have shallow potential wells, the 
energy deposition can have consequences that are dramatic in comparison
to normal galaxies. Several recent studies of the kinematics of
the HI gas in nearby galaxies (e.g. Puche
\& Westphal 1993, Young \& Lo 1996, 
Hunter et al. 1998, Wilcots \& Miller 1998, Van Zee, Skillman \& Salzer 1998) 
show clear
evidence of a disturbed ISM, where low density regions, often
surrounded by denser shells, expanding at velocities of the order of
$15$~km~s$^{-1}$ are identified. 
The most natural interpretation of
these data implies that these are regions of mechanical energy
injection due to stellar (winds and/or SNe) activity. The HI data can 
even be interpreted as a hint   that
the least massive galaxies can lose their gas, 
in a process that we will refer to in this paper
as blowaway. Indirect evidence that such a phenomenon might indeed
occur is given by Suntzeff \etal (1993). By studying the metallicity
distribution of stars in the Sextans dwarf galaxy, these authors find
that a possible explanation of the observed sharp cut-off  at high
metallicity is an an almost closed-box evolution followed by a quick
removal of gas, or, in our terminology, a blowaway.  

If the gas is
blown away from the galaxy, one might still expect to find it in the
surroundings of dSphs. 
There is currently no unambiguous
evidence for this (e.g. Bowen \etal 1997), although there are
tantalising suggestions, (e.g. Carignan \etal. 1998), but
the limits on diffuse, extended and/or hot gas are not very stringent.
ASCA
observations (Della Ceca \etal 1996; see also Heckman \etal 1995) of
the star-forming dwarf galaxy NGC1569 have been able to assess the
presence of diffuse hot gas ($T\sim 0.7$~keV) very likely heated by
supernova explosions.  Similarly, Bomans, Chu \& Hopp (1997) found
X-ray emission from hot gas within a supergiant shell in the dwarf
irregular NGC~4449, which also has a large-scale radio synchrotron
halo (Klein \etal 1996).  If the gas remains bound to the galaxy
rather than being expelled into the intergalactic medium, after a
cooling time it can rain back again onto the galaxy and become
available to refuel subsequent star formation, in a process
reminiscent of the ``galactic fountain'' thought to be at work in
larger spirals such as the Milky Way.  This circulation could in 
principle provide an attractive explanation for the intermittent star
formation activity in dSphs, such as Carina (Smecker-Hane \etal 1994) which
shows at least four distinct main sequence turn-offs separated by a
few Gyr. The difficulty with this scenario is that the required
residence time of the gas in the halo seems to be larger than its
typical cooling time, unless its density is very low, which is not
completely implausible. There are also liable to significant, difficult
to quantify, effects from nearby large galaxies.

If energy injection is mainly regulated by massive stars, then
the observed metallicity range of
dwarfs must be consistent with what is inferred from massive star
energetics and dust-to-gas ratio.
Several studies (Matteucci \& Tosi 1985, Clayton \& Pantelaki 1993, Marconi
\etal 1994; Lisenfeld \& Ferrara 1998) 
have already addressed this issue, and a consensus has been reached 
that all dwarf galaxies must have lost varying
amounts of gas and metals in the past, and probably that they continue
to do so today. 
Metal enhanced outflows
can explain the lack of prominent nuclei and the lower-than-average
metallicity of low mass systems (Larson 1974), and seem to
explain the luminosity-metallicity relationship for dIs (Lequeux
\etal 1979). The selective loss of metals, particularly oxygen, would
also resolve the discrepancy between the yields calculated from
stellar evolution theory and those measured by   observations (Pagel
\etal 1992; Maeder 1992).  
Dekel \& Silk (1986) 
concluded that in order to reproduce the surface brightness and
metallicity decline towards low masses, a significant gas loss in a
dominant DM          halo potential has to take place, with an upper
critical halo circular velocity of $\sim 100$~km~s$^{-1}$. 
Vader (1986, 1987) put the idea of "metal
enhanced" winds on a more quantitative basis.
Recent numerical experiments (MacLow \& Ferrara 1999, hereafter MF)
have shown that indeed during multi-supernova events 
in dwarf galaxies, most of the metals mixed with the hot cavity gas
are able to leave the galaxy, whereas the fraction of cool ISM 
gas lost is relatively small (at most $\sim$7\% of the galaxy mass). 
Nevertheless, if mixing of the two components in the wind, 
or with the hot, shell-evaporated interior takes place, 
this amount of gas is sufficient to dilute the wind metallicity
to a level comparable or only slightly higher
than the one of the ISM of the parent galaxy.   
We will discuss this point in more detail in the section 2.1.1.

Dark matter is the second important ingredient necessary
to properly understand dwarf galaxy evolution.
Data on the DM content of low mass galaxies, although slowly
accumulating are still scarce, and often difficult to interpret.
The $M/L$ ratios have been shown to vary between 5.7 
and 94 (e.g. Mateo \etal 1993, Pryor 1996), confirming the 
perception that small objects tend to be dark
matter dominated. The same conclusion can be reached from rotation
curve studies: although exploring the mass distribution only
down to relatively large values, a clear trend of increasing dark-to-visible
mass ratios with decreasing mass is seen (Mateo 1993). 

In this study we consider
the effects of stellar feedback and DM         
on the evolution of dwarfs from a new perspective.  There have been a
few attempts to combine chemical and dynamical evolution models for
dwarf galaxies of various types (Koeppen \etal 1995; De Young \&
Heckman 1994; De Young \& Gallagher 1990; Burkert, 
Theis \& Hensler 1993; Silk 1997).  Our approach follows
naturally from the evidence that supernovae  (SNe) and
superbubbles (SBs) are the main contributors to the ISM energy
budget.  Metals are
injected into the ISM during every SN explosion event, and if an
outflow develops, 
it allows gas to escape from the disk and even from the
gravitational potential of the DM          halo. The amount of gas and
metals lost is regulated by the pressure in the ISM,
which in turn depends upon the energy injection by previous
SNe. The pressure is increased as the result of the 
fact that gas random motions induced by SN explosions 
generate a non-thermal
(turbulent) pressure in addition to the thermal one.
Ultimately the pressure can be increased up to the point at which bubbles
produced by SNe are confined in the main body of the galaxy, thus
inhibiting the outflow. Depressurisation occurs as the parcels of gas
(clouds) move and
collide at supersonic velocities with
respect to their internal sound speed; radiative dissipation behind 
shocks following collisions then stabilizes the
system. On this basis, we expect a correlation between overall
dynamical properties, such as gas velocity dispersion, and metallicity
of the galaxy.

Very few detailed observations of the velocity
dispersion of HI gas in dIs have been made; the situation is made worse
by the lack of abundance measurements for those galaxies for which
velocity information is available.
Only for a handful of galaxies 
velocity dispersion and metallicity are simulatneously available and so 
no correlation
has yet been firmly established.  However, some studies 
(Lequeux \etal 1979; Hunter \etal 1982; 
Matteucci \& Chiosi 1983; Skillman, Kennicutt \& Hodge 1989) suggest
a general trend of increasing metallicity with
dynamical mass.  

The main aim of this paper is to include the above physical processes,
governed by stellar feedback and DM         , in a consistent
evolutionary model of dwarf galaxies able to explain their most
fundamental properties.

The plan of the paper is as follows. In Sec. 2 we present the basic
assumptions and equations of our model; Sec. 3 is devoted to 
simple, but nevertheless physically significant, analytical insights.  
Sec. 4 discusses the available observations and the dI galaxy sample 
to which we compare our results.
In Sec. 5 we present the results and in Sec. 6 we discuss them along
with a possible evolutionary scenario.
A brief summary of the results in Sec. 7 concludes the paper.

\section{Basic Equations}

The equations describing the rate of change of {\it (i)} the
galactic gas mass, $M_g$; {\it (ii)} the mass fraction of a given
heavy element $i$, denoted by $X_i=M_i/M_g$ ($M_i$ is the mass of the
element $i$); {\it (iii)} the kinetic energy stored in ISM bulk
motions, $\epsilon_k=\rho \sigma^2$, where $\rho$ is the average ISM density 
and
$\sigma$ its turbulent velocity dispersion, can be written as
\begin{equation}
\label{mg}
{d\over dt}M_g(t)= -\psi(t) + E(t) + A(t) - W(t),
\end{equation}
\begin{equation}
\label{xi}
{d\over dt}[X_i(t)M_g(t)]= -X_i(t)[\psi(t)+W(t)] + E_i(t) +
X_{inf}A(t),
\end{equation}
\begin{equation}
\label{sig}
{d\over dt}[\rho \sigma^2(t)]= \dot\epsilon(t)^+-\dot\epsilon(t)^-.
\end{equation}
We will devote each of the following three subsections to the detailed
discussion of the various terms appearing in eqs. \ref{mg}-\ref{sig}.
Before doing that, though, it is necessary to describe our galaxy
model.

For our purposes, we model a dwarf galaxy as a system initially made of a
visible disk with combined gaseous and stellar mass $M_g$, and a DM
halo of mass $M_h$.
We assume that the gas has a density distribution given by
$\rho(\varpi,z,\phi)=\rho_0 f(\varpi,z)$, where $\rho_0=\rho(0,0)$,
$\varpi$ is the galactocentric radius, $z$ is the vertical coordinate,
and $f(\varpi,z)$ is a function to be determined, for example, by
imposing hydrostatic equilibrium of the gas in the total gravitational
potential of the dwarf galaxy,
$\Phi_t(\varpi,z)=\Phi_h(\varpi,z)+\Phi_d(\varpi,z),$ as due to the
halo and disk components.

The behavior of the dark-to-visible mass ratio $\phi=M_{h}/M_g$ in
galaxies has been explored in great detail in a key study by Persic
\etal (1996, PSS). These authors find that $\phi$ is a function of the
galactic mass; using their relations one can derive the
dependence of this ratio on the visible mass of the galaxy:
\begin{equation}
\label{pss}
\phi \simeq  34.7 M_{g,7}^{-0.29}.
\end{equation}
where $M_{g,7}=M_g/10^7~M_\odot$. From this equation it is clear that
the gravitational potential of dwarf galaxies with $M_g \lesssim 10^9
M_\odot$ is dominated by the DM halo; therefore we can
neglect the potential due to visible mass $\Phi_d$. In the rest of
the paper we will not make any specific assumption on $\phi$, apart
from specific examples where we will use eq. \ref{pss}.

The density profiles of dwarf DM halos remain uncertain.  A comparison
between CDM cosmological model predictions and the available
observational data, discussed by MF, shows that
a considerable disagreement still exist.  In view of these
uncertainties, we calculate the halo gravitational potential by
assuming that the density distribution of the halo can be approximated
by a modified isothermal sphere (Binney \& Tremaine 1987), which is
general enough to be appropriate for an idealized situation such as
the one presented here, and does reproduce the observed central core.
It follows that
\begin{equation}
\label{rhoh}
\rho_h(r)= {\rho_c \over 1+ (r/r_0)^2}.
\end{equation}
The halo mass as a function of radius is then
\begin{equation}
\label{mh}
M_h = M_h(r_h) \simeq 4\pi\rho_c r_0^2 r_h,
\end{equation}
where $r_h$ is an appropriately defined halo radius. Following a
common {\em Ansatz} we take
\begin{equation}
\label{rh}
r_h\equiv r_{200}= \left(\frac{3\rho_c}{200 \rho_{crit}}\right)^{1/2}
r_0,
\end{equation}
where $r_{200}$ is the characteristic radius within which the mean DM
density is 200 times the present critical density, $\rho_{crit}=3
H_0^2\Omega/8\pi G=1.88\times 10^{-29} h^{-2} $~g~cm$^{-3}$, where
$H_0= 100 h$~km~s$^{-1}$~Mpc$^{-1}$ is the Hubble constant; throughout
the paper we will use $\Omega=1$ and $h=0.5$.  To proceed further we need a
relation between the scale radius, $r_0$, the central DM density
$\rho_c$, and the mass of the halo $M_h$.  Burkert (1995) has shown
that the total DM inside $r_0$, given by $M_0=M_h(r_0) =
\pi(4-\pi)\rho_c r_0^3$, is related to $r_0$ and $\rho_c$ through the
relations
\begin{equation}
\label{grels}
M_0= 7.2\times 10^7 \left({r_0\over {\rm
kpc}}\right)^{7/3}~~M_\odot,~~~~~~~~~
\rho_c= 2.7\times 10^7 \left({r_0\over {\rm kpc}}\right)^{-2/3}~~M_\odot~{\rm
kpc}^{-3}.
\end{equation}
Substituting $M_0$ and $r_h$ from eq. \ref{rh} into
equations~(\ref{grels}) we get
\begin{eqnarray}
\label{r01}
r_0 & = &5.3\times 10^{-5} \left({M_h\over M_\odot}\right)^{1/2}
h^{1/2} \mbox{ kpc}, \\
\label{rhoc1}
\rho_c &=& 2\times 10^{10} \left({M_h\over M_\odot}\right)^{-1/3}
h^{-1/3} M_\odot \mbox{ kpc}^{-3}.
\end{eqnarray}
With these assumptions the gravitational potential of the halo is
\begin{equation}
\label{grav}
\Phi_h(r) = 4\pi G \rho_c r_0^2 \left[{1\over 2} \log \left(1+ x^2\right) + 
{\arctan x\over x}\right],
\end{equation}
where $x=r/r_0$. The galaxy has a circular velocity
\begin{equation}
\label{vc}
v_c^2(r) = r {\partial \Phi_h\over\partial r}= {4\pi G\rho_c
r_0^2\over x} (x-\arctan x);
\end{equation}
The rotation curve increases rapidly in the inner parts of the galaxy,
already being practically flat at $x=1.5$.

In analogy with the DM component, we suppose that the gas extends out
to a disk cut-off radius, $\varpi_*$.  From a fit to the sample of
dIs (discussed in Sec. 4) that we will use to test the model, it has
been found that the HI radius-mass relation is well approximated by
the law
\begin{equation}
\label{rc}
\varpi_* =  \varpi_0 M_{g,7}^\alpha\simeq 1.5  M_{g,7}^{0.338}~~~ {\rm
kpc},
\end{equation}
where $M_{g,7} = M_g/10^7 M_\odot $.
To obtain this relation we have in addition assumed that the ratio
between the HI and the optical radius is $\approx$ 2. This is 
roughly consistent with the results of Salpeter \& Hoffman (1996) 
who found a value $2.34\pm 0.14$.
We will use eq. \ref{rc} and the value $\varpi_0=1.5$~kpc throughout the 
paper, 
even if some uncertainty may still be present, as discussed
in Sec. 6.  For sake of clarity, it might be
useful to give the explicit expressions for the relations among the
three characteristic radii so far introduced in this Section. These
are:
\begin{eqnarray}
\label{rrels}
{r_h \over \varpi_*} &\simeq& 1.23 (\phi h^{-2})^{1/3}\\ {\varpi_*
\over r_0} &\simeq& 15.6 M_{g,7}^{-1/6}(\phi h)^{-1/2}\\ {r_h \over
r_0} &\simeq& 19.2 M_{g,7}^{-1/6}(\phi h^{7})^{-1/6}.
\end{eqnarray}
In order to provide convenient expressions for the gas 
scale height $H$, number density $n_0=\rho_0/\mu m_h$, and
column density $N_{H}$, we assume that most of the mass is
concentrated in a thin disk, producing a constant gravitational
acceleration $g= 2\pi G
\Sigma_t$, where $\Sigma_t=(M_g+M_h)/2\pi \varpi_*^2$ is the total
matter surface density. Such a choice generates an exponential gas
distribution with parameters given by the following formulae:
\begin{eqnarray}
\label{nhn}
H&=& {c_{s,eff}^2\over 2\pi G \Sigma_t}={c_{s,eff}^2 \varpi^2\over G
M_g(1+\phi)}\simeq 2 \left({\varpi_0^2 \over \phi}\right)
M_{g,7}^{2\alpha-1} c_{10}^{2}~~~{\rm kpc}, \\
\label{n0}
n_0&= &{M_g\over 2\pi H \varpi^2 \mu m_h}\simeq 2\times 10^{-2}
\left({\phi\over \varpi_0^4}\right) M_{g,7}^{2(1-2\alpha)}
c_{10}^{-2}~~~{\rm cm}^{-3}, \\ N_H &=& n_0 H \simeq 2\times 10^{20}
M_{g,7}^{(1-2\alpha)}
\varpi_0^{-2}~~~{\rm cm}^{-2}, 
\end{eqnarray}
where $c_{10}=c_{s,eff} /{\rm 10~km~s^{-1}}$, $c_{s,eff}^2=c_{s}^2 +
\sigma^2$ is an effective sound speed, $c_s$ is the gas sound speed;
$\varpi_0$ is defined by
equation~(\ref{rc}) in kpc, and $\mu$ is the mean molecular weight.
We have also assumed that $\phi\simgt 1$ to simplify the last 
expressions.
We note that in this approximation, $N_H$ is independent of
both $\phi$ and $c_{s,eff}$ and, given that $1-2\alpha \simeq 1/3$,
only weakly dependent on the visible mass of the galaxy.

\subsection{Gas Mass}

Eq. \ref{mg} describes the evolution of the galactic gas mass
$M_g$. The gas content is decreased by star formation and outflow
phenomena, at rates $\psi(t)$ and $W(t)$, respectively; it is
increased by infall and stellar ejecta, at rates $A(t)$ and $E(t)$,
respectively.

The star formation rate $\psi$ has been suggested by different authors
to depend on various quantities as the gas mass, column density or
number density.  We choose $\psi$, following Dekel \& Silk (1987), to
be proportional to the gas mass divided by the galactic free-fall time
$t_{ff} = (4\pi G \rho)^{-1/2}$, where $G$ is the gravitational
constant: $\psi= M_g/\tau t_{ff}$; $\tau \simeq 160$ is the value
appropriate to reproduce the actual star formation rate in the Milky
Way. In addition to its simplicity, this formula represents the most
natural scaling for the process and it can be physically motivated.
We will see below as this expression results is a Schmidt-type law,
modified by the presence of DM.  The gas mass returned from stars,
$E(t)$, depends on parameters as the stellar IMF, yield and return
fraction, which are defined and discussed in Sec. 2.2.

The infall rate, $A(t)$, is very uncertain;
for this reason most of our calculations assume no
infall ($A(t)=0$). We therefore presume that 
infall should not represent a major
effect for low mass galaxies.

The term describing the gas outflow rate $W$ is complex, and it is
discussed in the following section, where we also introduce the distinction
between blowout and blowaway processes.

\subsubsection{Outflows}

\centerline{\it (a) Blowout}

Both SNe and SBs may contribute to drive a substantial mass loss from
the galaxy, but it is likely that SBs are more efficient due to their
higher energy input; in the following we will mainly describe the SB case, 
but similar treatment applies to SNe.  In order that mass previously located
inside the region affected by the (multiple) explosion is ejected from
the galaxy, the velocity $v_b$ of the expanding
shell has to be 
larger than the escape velocity from the galaxy, $v_e$. The
escape velocity can be calculated at the disk radius $\varpi_*$ from
the potential eq. \ref{grav}:
\begin{equation}
\label{ve}
v_e^2(\varpi_*) = 2 \vert \Phi_h(\varpi_*) \vert \sim 8\pi G \rho_c
r_0^2 \left[{1\over 2}
\log (1+ x_*^2) + {\arctan x_*\over x_*}\right]; ~~~x_*=\varpi_*/r_0.
\end{equation}
Taking the appropriate value for $x_*$, as derived from
eq. \ref{rrels}, the terms in parenthesis give a factor $\simeq 1.65$
for all galactic gas masses, since it is $x_* \gg 1$. It follows that
\begin{equation}
\label{ve1}
v_e(\varpi_*) \sim (13.2\pi G \rho_c)^{1/2} r_0 \sim 20
M_{g,7}^{1/3}(\phi h)^{1/3}~~~{\rm km~s}^{-1},
\end{equation}
Note that $v_e$ is basically independent of $\varpi_*$ as long as
$x_*\gg 1$, which implies an almost flat $v_e(\varpi)$.

The evolution of a point explosion in an exponentially stratified
medium, as the one introduced above, has been derived by
Kompaneets (1960). This solution accurately approximates the
exact numerical solution, (see for example MacLow \etal 1989) and can
be obtained from dimensional analysis. It is useful to rederive it
briefly in this context.

Suppose that the gas density distribution is horizontally homogeneous
and $\rho(z)=\rho_0 \exp (-z/h)$.  The velocity of the shock wave is
$v\sim (p/\rho)^{1/2}$, where the pressure $p$ is roughly equal to
$E/z^3$; $E=N E_0$ is the total energy of the explosion, produced by
$N$ SNe in the association (we usually take $N\approx 100$), each of
energy $E_0=10^{51}$~ergs.  Then, it follows that
\begin{equation}
\label{v}
v(z) \simeq E^{1/2} \rho_0^{-1/2} e^{z/2h} z^{-3/2}.
\end{equation}
This curve has a minimum at $z=3H$ and this defines the height at
which the shock wave, initially decelerating, is accelerated to
infinity and a {\it blowout} takes place. Therefore we will use
$3H$ as the fiducial height where the velocity $v_b=v(3H)$ is
evaluated.  Introducing the mechanical luminosity of the explosion,
$L=E/t_{_{OB}}$, where $t_{_{OB}} \simeq 3 \times 10^7$~yr is the
average lifetime of massive stars, we write eq. \ref{v} as
\begin{equation}
\label{vb}
v_b \simeq \left({e\over 3}\right)^{3/2} L^{1/2} \rho_0^{-1/2}
t_{3H}^{1/2} H^{-3/2},
\end{equation}
where $t_{3H}$ is the time at which $z=3H$. Unfortunately, the
Kompaneets solution does not allow to estimate $t_{3H}$ in a simple
manner.  For our purposes it is accurate enough to determine
$t_{3H}$ from solution for $v(t)$ given by Abbot \etal (1981) valid for a
homogeneous atmosphere,
\begin{equation}
\label{vt}
v(t) = {3\over 5} \left({125\over 154 \pi}\right)^{1/2} L^{1/5}
\rho_0^{-1/5} t^{-2/5},
\end{equation}
from which follows
\begin{equation}
\label{t3h}
t_{3H} = (3H)^{5/3} \left({125\over 154 \pi}\right)^{-1/3}
\left({L\over \rho_0}\right)^{-1/3}.
\end{equation}
The final expression for $v_b$, obtained by substituting $t_{3H}$
in eq.\ref{vb}, is then
\begin{equation}
\label{vb1}
v_b = {e^{3/2}\over 3^{2/3}} \left({125\over 154 \pi}\right)^{-1/6}
\left({L\over \rho_0}\right)^{1/3} H^{-2/3}= 92  L_{38}^{1/3}(1+\phi)^{1/3}
c_{10}^{-2/3}~~~{\rm km~s}^{-1},
\end{equation}
where $L_{38}=L/10^{38}$~ergs~s$^{-1}$ and we have used
eqs. \ref{nhn}-\ref{n0}.  Eq. \ref{vb1} shows that the velocity at the
re-acceleration point $z=3H$ is inversely proportional to the scale
height of the gas to the 2/3 power. As a consequence, thick gas layers
will be able to prevent blowout. Note that $H$ depends on
$\sigma$ (see eq. \ref{nhn}), in turn regulated by the SN energy
input, as described by eq. \ref{sig}.

There are three different possible fates for the
SN-shocked gas, depending on the value of $v_b$. If $v_b^2
\le c_{s,eff}^2$, where $c_s$ is the sound speed in the ISM, 
the explosion will be confined in the disk and no mass loss will
occur; for $c_{s,eff}^2 < v_b^2 < v_e^2$ the supershell will breakout
of the disk into the halo, but the flow will remain bound to the
galaxy; finally, $v_b > v_e$ will lead to a true mass loss from the
galaxy. In the second case, it is likely that the hot gas will fall
down again onto the disk, in a time scale comparable to the radiative
cooling time
\begin{equation}
\label{tauc}
\tau_c \sim 3.2\times 10^7 \left({T\over 10^6 {\rm K}}\right)^{1.6}
\left({n\over 10^{-3} {\rm cm}^{-3}}\right)^{-1}~~{\rm yr},
\end{equation}
where $T$ and $n$ are the gas temperature and density,
respectively. Since $\tau_c$ is much shorter than the typical chemical
timescale ($\simgt 10^9$~yr), the delay on which the infall takes
place with respect to the outflow is negligible, and to all purposes
we can assume that the halo gas has always remained in the disk.

Once the blowout velocity $v_b$ has been determined, the mass outflow
rate can then be written as
\begin{equation}
\label{w}
W=\cases{0 & if $v_b < v_e$\cr 2\xi E_k^{(j)}
\gamma^{(j)} v_b^{-2} & if $v_b > v_e$} 
\end{equation}
Note the inverse dependence on $v_b^2$ of $W$: this simply reflects the
fact that for a given driving energy $\propto \rho v_b^2$, slower outflows 
are more mass loaded.
In the previous expression the kinetic energy injected by the source
$j$ (SNe or SBs) is $E_k^{(j)}=\eta E^{(j)}$, where $\eta_j$ is an
appropriate efficiency coefficient, that can be estimated from the
assumed expansion law. For example, the expansion law given above has
$\eta=3/7$. However, for radiative bubbles, $\eta$ is smaller and 
equal to about 3\% (Koo \& McKee 1992); we use this value throughout the paper.

Since part of the kinetic energy is used to accelerate material in the
equatorial plane of the bubble, only a mass fraction $\xi $ will leave
the galaxy in the blowout. The actual value of $\xi$ is relatively
uncertain and model dependent. In a recent study MF 
have calculated, by means of a series of numerical simulations
that assume a galactic structure identical to the one adopted here,
the value of $\xi$ in the range of mechanical luminosities
$L=10^{37-39}$~ergs~s$^{-1}$ and galactic masses $M_{g,7}=0.1, 100$.
They find that in general the gas ejection efficiency in case of
blowout is relatively small ($\xi \simlt 7$\%). This conclusion
has been obtained by keeping $L_{38}$ fixed throughout the entire energy
injection phase, and, more crucially, assuming a power-law dependence
of $\phi$ on the mass of the galaxy given by PSS (eq. \ref{pss}). 
Since here we try not to make any assumption on such relation,
we choose to adopt the value $\xi = 0.07$, which should provide a
reasonable estimate of the mass ejection fraction.  

Finally, $\gamma^{(j)}$ is the explosion rate for the source $i$. For SBs it is
$\gamma^{(SB)} = S(t)f_{_{OB}}/N_{SN}$, where $S(t)$ is the SN rate in
the galaxy and $f_{_{OB}}$ is the fraction of SNe occurring in an
association ($f_{_{OB}}\sim 0.7$). The analogous quantity for isolated
SNe is $\gamma^{(SN)} = S(t)(1-f_{_{OB}})$.  The SN rate is obtained
from $S=\nu \psi$, where $\nu$ is the number of SNe per unit mass of
stars formed which can be calculated once a given IMF has been
specified (see Sec. 2.2). 
%

\centerline{\it (b) Blowaway}

By definition, the blowout phenomenon discussed above involves a
limited fraction of the parent galaxy mass, the one contained in 
cavites created by the SN/SB explosion. A much more disruptive event
may occur, \ie the blowaway, in which the gas content of the galaxy is
completely expelled and lost to the intergalactic medium. 
In the following we derive in
detail the necessary condition for the blowaway to occur.

We have seen (eq. \ref{v}-\ref{vb}) that blowout takes place when the
at $z=3H$ the shell velocity exceeds the escape velocity.  Following
the blowout the pressure inside the cavity drops suddenly due to the
inward propagation of a rarefaction wave.  The lateral walls of the
shell, moving along the galaxy major axis will continue to expand
unperturbed until they are overcome by the rarefaction wave at
$\varpi=\varpi_c$, corresponding to a time $t_c$ elapsed from the
blowout.  After that moment, the shell enters the momentum-conserving
phase, since the driving pressure has been dissipated by the blowout.
The correct determination of $t_c$ requires the solution of a 5-th
order algebraic equation (De Young \& Heckman 1994).  In an
exponential atmosphere the shock wave moves faster in the vertical
direction than in the equatorial one; hence the ratio of the vertical
to equatorial velocity is $v_z(z)/v_\varpi = \exp(z/3H)$ and, when
blowout takes place, $v_b=\exp(3/2) v_{\varpi_b}$. At the same moment,
the radius of the shell in the plane is $\varpi_b=(2/3)H=b H$; the
rarefaction wave is travelling at the sound speed of the hot cavity
gas, $c_{s,hot}$.  It follows from simple kinematic considerations
that
\begin{equation}
\varpi_c \sim \varpi_b + {v_{\varpi_b}\over c_{s,hot}}(3H + \varpi_c )=a H,
\end{equation}
with $a$ to be determined.

\begin{figure}
\centerline{\psfig{figure=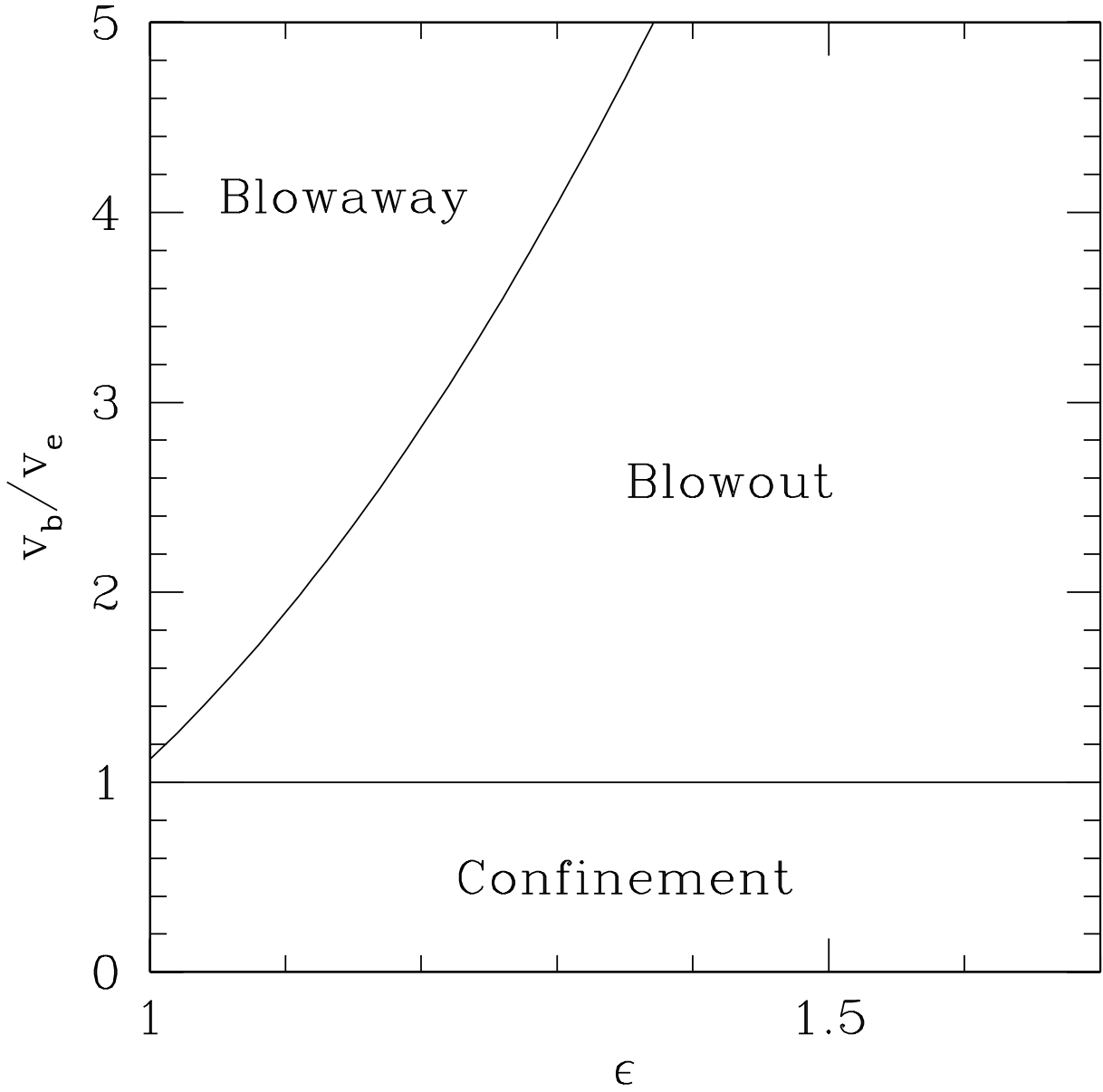}}
\caption{\label{fig1}\footnotesize{Conditions for blowaway, blowout and 
confinement as
a function of the major-to-minor axis ratio $\epsilon=\varpi_*/H$
of dwarfs galaxies; $\epsilon=1$ corresponds to spherical bodies.}}
\end{figure}

The requirement for the blowaway to take place is that the momentum of
the shell (of mass $M_c$ at $\varpi_c$) is larger than the momentum
necessary to accelerate the gas outside $\varpi_c$ (of mass $M_o$)at a
velocity larger than the escape velocity:
\begin{equation}
\label{mom}
M_c v_{\varpi_c} \ge M_o v_e,
\end{equation}
where $M_c = M_w + \pi(\varpi_c - \varpi_b)^2 \rho_0 H [1-(1/e)]$,
$v_{\varpi_c}=v_b e^{-3/2}(\varpi_b/\varpi_c)^{3/2}$, $M_o=
\pi(\varpi - \varpi_c)^2 \rho_0 H [1-(1/e)]$. Defining the axis ratio 
as $\epsilon=\varpi_*/H\ge 1$, substitution into eq. \ref{mom} yields
the blowaway condition:
\begin{equation}
\label{vbve}
{v_b\over v_e}\ge {(\epsilon-a)^2 e^{3/2}\over
\left[(1-\xi)a^2+(a-b)^2\right]}{\left(
a\over b\right)}^{3/2}.
\end{equation}
However, as $v_{\varpi_b}/c_{s,hot} \ll 1$ (see eq. \ref{vb1}), it
follows that assuming ${\varpi_c}\simeq {\varpi_b}$, or $a\simeq 2/3$,
is an excellent approximation for our purposes. Eq. \ref{vbve} then
becomes
\begin{equation}
\label{cond}
{v_b\over v_e}\simgt (\epsilon - a)^2 a^{-2} e^{3/2},
\end{equation}
if $\xi \ll 1$ as discussed above; eq. \ref{cond} is graphically
displayed in Fig. \ref{fig1}.  Flatter galaxies (large $\epsilon$
values) preferentially undergo blowout, whereas rounder ones are more
likely to be blown-away; as $v_b/v_e$ is increased the critical value
of $\epsilon$ increases accordingly. Unless the galaxy is perfectly
spherical, blowaway is always preceded by blowout; between the two
events the aspect of the galaxy may look extremely perturbed, with one
or more huge cavities left after blowout.
 
\subsection{Chemical Evolution}

Let us now consider the various terms appearing in eq. \ref{xi}.  As
stated above the gas mass returned from stars, $E(t)$, is dependent on
the IMF, $\varphi$, the return fraction, $R$ (the fraction of the
total mass transformed into stars which is subsequently 
ejected back into the ISM), and the net stellar yield, $y$ (the total mass
of an element ejected by all stars back into the ISM per unit mass
of matter locked into stars). The IMF is of critical importance, since
it determines the relative number of high mass (which eventually will end
their life as Type II SNe) and low mass stars. However, its true
nature is not well understood. Usually a single power-law function is
used (there is little compelling evidence for a multi-component fit
[Mateo 1988]; see also Padoan \etal 1997) with small
differences among the 
most frequently used indexes. We follow this convention and assume 
a power-law form of the IMF, $m\varphi(m)\propto m^{-x}$, with $x$
given below.

We next adopt the {\it Instantaneous Recycling Approximation}, (IRA)
\ie stars less massive than 1 $M_\odot$ live forever and the others
die instantaneously (Tinsley 1980), which allows $E$ and $E_i$ to be
written respectively as
\begin{equation}
E=R\psi;~~~~~~~~~~~~~~~~~~~~~~~ E_i=[RX_i + y(1-R)]\psi;
\end{equation}
note that Tinsley's eq. 3.14 contains a wrong extra factor $1-X_i$,
see Maeder (1992). The limitations of IRA are discussed by Tinsley
(1980).

In order to quantify the above relation, we need to fix the value of
$R$ and $y$ for the traced heavy element. We choose oxygen for the
following reasons: {\it (i)} it is produced mainly in Type II SNe, which are
also governing the mass loss and dynamics of the galactic ISM via
their energy input; {\it (ii)} a large sample of dwarf galaxies with good
quality abundance data for this species is available (see Sect. 4);
{\it (iii)} oxygen yields have also been extensively studied by several
authors (Arnett 1978, Woosley \& Weaver 1986, Thielemann, Nomoto \&
Hashimoto 1992, Maeder 1992) and there is a good consensus about their
dependence on the stellar mass. 
We
therefore take $X_i\equiv X_o$, the oxygen mass fraction.  The
returned fraction and the net yield can be obtained using the standard
formulae (Maeder 1992):
\begin{equation}
\label{yields}
R = {\int_{m_l}^{m_u} (m - w_m) \varphi (m) dm \over
\int_{m_l}^{m_u} m \varphi (m) dm};\qquad  y = {1\over (1-R)}{\int_{m_l}^{m_u}
m p_m \varphi (m) dm \over
\int_{m_l}^{m_u} m \varphi (m) dm}
\end{equation}
where the IMF $\varphi$ is defined between the lower and upper masses
$m_l=1 M_\odot$ and $m_u=120 M_\odot$, and $w_m$ ($w_m=0.7 M_\odot$
for $m\le 4 M_\odot$ and $w_m=1.4 M_\odot$ for $m > 4 M_\odot$) is the
remnant mass (white dwarf or neutron star). The value of the power law
index is $x=1.35$ (standard Salpeter) and $x=1.7$ (Scalo 1986).  We
have taken the oxygen stellar yield, $p_m$ (\ie the mass fraction of a
star of mass $m$ converted into oxygen and ejected) from Arnett
(1990). With these assumptions we obtain $R=(0.79, 0.684)$ and
$y=(0.0871, 0.0305)$ for the (Salpeter, Scalo) IMF, respectively. We
will present results only for the Salpeter IMF (although we have
studied a few cases with the Scalo IMF) as differences are relatively
minor and Salpeter's law seems to be currently better supported
by a variety of observational evidences. 
 
The previous values for the yields seem to be robust, although
recent studies have pointed out that they might somewhat vary with 
metallicity. For example, using new stellar evolution models 
for $Z\approx 0.001$ 
(1/20$^{\rm th}$ solar) and $Z\approx 0.02$ (solar), which include new 
opacities, Maeder (1992) calculated stellar yields for, among
other species, oxygen as a function of metallicity and the expected mass limit
for the creation of black holes. The stellar models take into account metal
dependent opacity and nuclear effects, changes of mass loss rate with
metallicity and moderate core-overshooting. 
Yields of heavy elements tend to decrease with stellar mass if a  
black hole is formed, because in this case heavy 
elements are partially swallowed by the black hole. This
effect is less important at  high metallicities, because most     metals 
escape in a stellar outflow before the collapse to a black hole. 
In order to include these findings, at least in an approximate manner,
we have used interpolated Maeder's results for case C (minimum mass for
formation of the black hole equal to 22 $M_\odot$) yields. For a Salpeter
IMF the value of the yield agrees with the one obtained from Arnett's
results at metallicity $\approx 15$\% of solar, but it is smaller for
lower metallicities. 
We have done numerical runs of our models using both prescriptions
and we found differences of the order of 30\% in the final metallicity.

A particularly important quantity depending on the IMF is the number
of SNe per unit mass of stars formed, $\nu$. For a Salpeter IMF with
$m_l=1 M_\odot$ and $m_u=120 M_\odot$, we obtain $\nu^{-1}=53
M_\odot$.  It is worth pointing out, though, that $\nu$ is rather
sensitive to the adopted value of $m_l$.

Finally, we are making the hypothesis that the oxygen abundance in the
outflow is the same as in the ISM, \ie we are not
considering metal-enhanced outflows. The reason for this assumption is
the following.  MF               have determined, in addition to the
gas ejection efficiency as discussed in the previous Section, the
metal ejection efficiency per SB in a dwarf galaxy, $\xi_Z$. Let us
consider a typical case among the ones simulated in their paper, \ie
$L_{38}=1$, $M_{g,7}=1$, for which they obtain $\xi = 1.4\times
10^{-2}$ and $\xi_Z = 1.0$. If the amount of heavy elements produced
per SN is $\approx 3 M_\odot$, then during the energy injection phase
lasting $50$~Myr, about $470 M_\odot$ of metals are diluted with the
ejected gas, if mixing between the two components is fast enough. This
corresponds to an increase in the metal abundance of the outflow of
$\approx 3\times 10^{-3}$, which is at least comparable if not
smaller, than the typical abundances of dIs, as can be appreciated by
inspecting our control sample in Tab. 1. This should demonstrate that
metal enhanced outflows 
can be produced only in the very early stages of the evolution.
In addition the above argument does not take into account the interaction
with the hot cavity gas which strengthens this conclusion.
Of course, if
all the metals produce in a SB escape from the galaxy as in the case
just considered, there would be no way for the galaxy to become metal
enriched.  However, the results of MF                       allow to
conclude that the metal enrichment is mostly due to low-luminosity SBs
and/or isolated SNe, which do not blowout and for which $\xi_Z$ is strongly reduced.

\subsection{Kinetic Energy Balance}

\begin{figure}    
\centerline{\psfig{figure=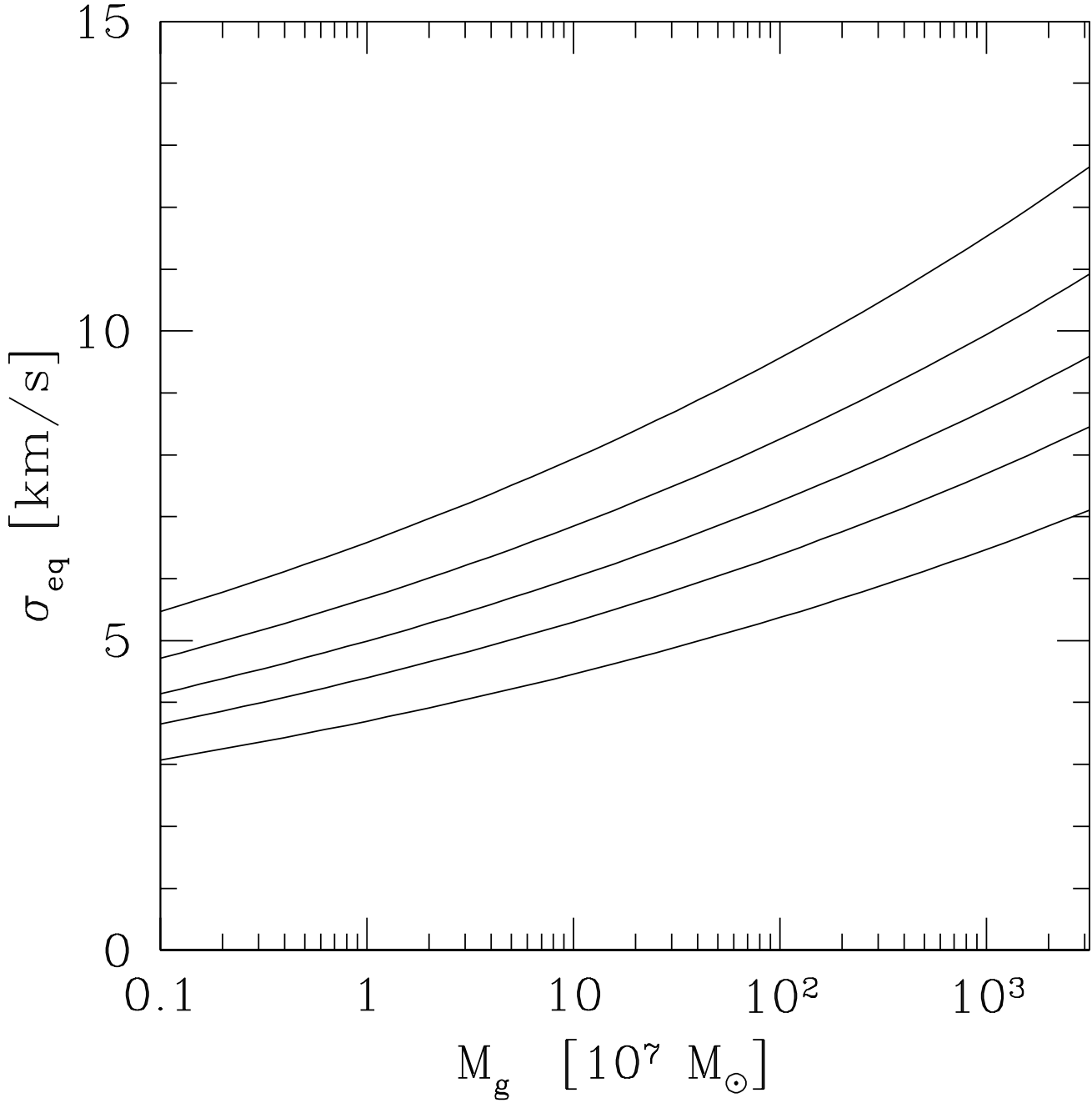}}
\caption{\label{fig2}\footnotesize{ Velocity dispersion equilibrium value, 
$\sigma_{eq}$, as 
a function of galactic gas mass and for different values of $\phi=0,
3, 10, 30, 100$ from the lowermost to uppermost curve, respectively.}}
\end{figure}

Finally, we consider the kinetic energy balance, eq. \ref{sig}. As we
have seen in previous Sections, SNe and SBs are responsible for
the mass outflow from the galaxy. On the other hand, they also
contribute to the turbulent pressure of the ISM. Turbulent motions
sustained by SN mechanical energy input (for a derivation of the
turbulent spectrum in the ISM, see Norman \& Ferrara 1996) can be
dissipated via radiative losses during cloud-cloud
collisions. Eq. \ref{sig} expresses the rate of change of the kinetic
energy density in the ISM, $\epsilon_k=\rho \sigma^2$, as determined
by the turbulent energy sources and sinks.  The rate at which
$\epsilon_k$ increases following SNe and SBs explosions is
\begin{equation}
\label{eplus}
\dot\epsilon^+ ={\rho\over M_g} \left[E_k^{(j)}\gamma^{(j)}(1-\xi)\right]= 
1.2\times 10^{-28}\left({\phi \over \varpi_0^4}\right)^{3/2}
M_{g,7}^{3(1-2\alpha)} (c_s^2 + \sigma^2)_{10}^{-3/2}.
\end{equation}
Note that eq. \ref{eplus} is strictly valid only for $\phi \simgt 1$,
due to the approximations made in eq. \ref{nhn}-\ref{n0}.
The dissipation rate per unit volume $\dot\epsilon^-$ is the product
of three quantities: the energy lost per collision, $\Delta E$, the
number density of clouds in the galaxy, $N_{c}$, and the cloud-cloud
collision frequency, $\omega_c$.  If $R_c \approx 5$~pc is the cloud
radius, then $N_c= 1/\pi R_c \lambda_{mfp}$, where
$\lambda_{mfp}\approx 0.13$~kpc is the mean free path for cloud-cloud
collisions; these values are conservatively assumed to be the same as
in the Milky Way (Spitzer 1978). Then, the collision frequency is
$\omega_c=\sigma/\lambda_{mfp}$.  In general, one should integrate the
dissipation rate over a cloud velocity distribution;
for simplicity, we assume that all 
clouds move with the same velocity dispersion $\sigma$.  The maximum
energy loss per collision $\Delta E$ takes place when the collision is
inelastic: for two clouds of mass $m_1$ and $m_2$ moving with relative
velocity $\Delta {\bf v}= {\bf v}_1-{\bf v_2}$ this corresponds to
\begin{equation}
\label{edelta}
\Delta E = {1\over 2}\vert \Delta v\vert^2 {m_1 m_2\over m_1+m_2}.
\end{equation}
Given the present uncertainty on the determination of the mass
spectrum of diffuse clouds in dwarf galaxies, we assume that all
clouds have the same mass, $m$.  It follows that $\Delta E=
m\sigma^2$, \ie all the initial kinetic energy is lost.  Are cloud
collisions in dwarfs truly inelastic ? The elasticity of collisions
(defined as the ratio of the final to the initial kinetic energy of
the clouds) has been recently studied by Ricotti \etal (1997).
Among
other results, the authors find that the elasticity depends both on
metallicity and on cloud size; nevertheless, most conditions are found
to lead to inelastic collisions, in the absence of a magnetic field.
In two subsequent papers, Miniati \etal (1997, 1999) pointed out that
even a relatively weak magnetic field makes collisions much more
elastic. In fact, if a magnetic field is present, a considerable
fraction of kinetic energy can be stored in the bending modes of the
field and returned to the clouds after the collision. The  result is
in a quasi-elastic collision with a much smaller net dissipation.
Including all the detailed physics they considered would greatly
complicate the present picture, let alone our ignorance about
magnetic fields in dwarf galaxies. To roughly take into account this
effect, we have adopted a reduction factor $b_1 \simlt 1$ in the
energy dissipation rate.  Hence it is
\begin{equation}
\label{eminus}
\dot\epsilon^- = b_1 N_c \omega_c \Delta E = 
2.1\times 10^{-28} b_1 \left({\phi \over \varpi_0^4}\right)
M_{g,7}^{2(1-2\alpha)}
\sigma^3_{10} 
(c_s^2 + \sigma^2)_{10}^{-1}.
\end{equation}
Fig. \ref{fig2} shows the velocity dispersion equilibrium value,
$\sigma_{eq}$ obtained by equating $\dot\epsilon^+$ to
$\dot\epsilon^-$ (for $b_1=1$) 
as a function of galactic gas mass and for different
values of $\phi$.  These curves represent the limiting case in which
$c_{s,eff}$ is dominated by turbulence, \ie $c_s$ is negligible with
respect to $\sigma$. Then the above inequality holds to a good
approximation for dwarfs galaxies and the equilibrium value for
$\sigma$ is given by
\begin{equation}
\label{eqsigma}
\sigma_{eq} \simeq 4.5 b_1^{-1/4} \left({\phi \over \varpi_0^4}\right)^{1/8} 
M_{g,7}^{(1-2\alpha)/4} 
{\rm km ~s}^{-1}.
\end{equation}
The equilibrium value of the cloud velocity dispersion increases 
with galactic gas mass as a result of the more vigorous
star formation activity occurring in those objects. 
Also, the spread introduced by the different DM          
content is about a factor 2 only for a given galactic mass. 
This allows us to conclude that the {\it gas velocity dispersion is
only mildly influenced by the underlying gravitational potential
of the galaxy}. Instead, the relatively similar values commonly
observed in galaxies greatly differing in mass, suggest that 
self-regulation is at work through the processes described in this
Section.  

It is instructive to calculate the time necessary to reach equilibrium, 
$t_{eq}=\sigma_{eq}^2\vert {d\sigma^2/dt}\vert^{-1}$, for the velocity
dispersion:
\begin{equation}
\label{eqtime1}
t_{eq}=10^7 \left({\phi \over \varpi_0^4}\right)^{-1/2}
M_{g,7}^{(2\alpha-1)} \sigma_{10}^2 (c_s^2 + \sigma^2)_{10} {\rm ~yr}.
\end{equation}
This short timescale implies that the memory of the initial gas
dynamical conditions is washed away very rapidly and, since (see
eq. \ref{xit} below), the chemical evolution timescale is much longer
than $t_{eq}$, using $\sigma_{eq}$ is an excellent approximation for
several purposes.

\section{Analytical Insights}

Before we present the numerical results in the next Section,
we discuss a few limiting cases that can provide some
physical insight into the problem.  We have seen that $t_{eq}$ is much
smaller than the chemical evolutionary timescale; this motivates the
approximation $d\epsilon_k/dt\simeq 0$.  Assuming in addition
$A(t)=0$, the system eqs. \ref{mg}-\ref{sig} is reduced to
\begin{equation}
\label{mg1}
{d\over dt}M_g(t)= -\psi(t) + E(t) - W(t),
\end{equation}
\begin{equation}
\label{xi1}
{d\over dt}[X_i(t)M_g(t)]= -X_i(t)\psi(t) + E_i(t) -X_i W(t),
\end{equation}
or, substituting the expressions given in the previous Section for the
various terms,
\begin{equation}
\label{mg2}
{d\over dt}M_g(t)= (R-1)\psi(t) - W(t),
\end{equation}
\begin{equation}
\label{xi2}
M_g{d\over dt}[X_i(t)]= -y(R-1)(1-X_i)\psi.
\end{equation}
From eq. \ref{xi2} one can see that the oxygen fraction evolution does
depend on the outflow rate only through the changes in $M_g$, in turn
regulated by the outflow.  Integration of eq. \ref{xi2} yields
\begin{equation}
\label{xit}
X_i(t)=1-[1-X_i(0)]e^{y(1-R)/\tau t_{ff}t}.
\end{equation}
The chemical evolution timescale is $t_e\simeq t_{ff}\tau/y(1-R)$, which
for the parameters adopted here is $\approx 9\times 10^3 t_{ff}\gg
t_{eq}$, which justifies our statement after eq. \ref{eqtime1}.  
By eliminating time
between eqs.  \ref{mg2}-\ref{xi2}, we can write
\begin{equation}
\label{mgt}
M_g{dX_i\over dM_g}[(R-1)\psi-W]= -y(R-1)(1-X_i)\psi,
\end{equation}
Note that, unlikely for $dX_i/dt$, $X_i(M_g)$ is a function of the
outflow rate $W$. Since $R-1$ is always negative, the l.h.s. cannot
vanish for any value of $W$.

\subsection{No Outflow case: $W=0$}

When $W=0$, eq. \ref{mgt} has the simple solution

\begin{equation}
\label{ximg}
X_i(M_g)=1-[1-X_i(0)]\left[{M_g\over M_g(0)}\right]^y.
\end{equation}
 
\subsection{Constant Outflow case: $W={\rm const.}\ne 0$}

The solution of eq. \ref{mgt} for the case $W\ne 0$, (but constant) is
\begin{equation}
\label{ximg1}
X_i(M_g)=1-[1-X_i(0)]\left[{KM_g-1\over KM_g(0)-1}\right]^y,
\end{equation}
where $K=(1-R)/t_*W$ and $t_*=\tau t_{ff}$.  Since $t_*W$ is the
amount of gas ejected during a star formation timescale, $KM_g$
expresses the ratio between the mass of gas that goes into stars and
the gas loss via the outflow. Hence large values of $\vert K\vert$
refer to negligible ejection rates, whereas for small $K$ outflow mass
loss dominates.  Of course, eq. \ref{ximg1} reduces to eq. \ref{ximg}
in the limit $\vert K\vert\rightarrow \infty$.
 
\subsection{Are Outflows Important ?}

The metallicity of a dwarf galaxy is closely related to the amount
of gas consumption; it is then important to assess the relative
importance of the two main processes involved in the latter, 
namely star formation
and outflows.  A straightforward answer can be obtained by comparing the
two r.h.s. terms in eq. \ref{mg2}.  Outflow mass losses will be
dominant if
\begin{equation}
\label{outcond}
(1-R)\psi \ll W = 2\xi E_k^{(j)} \gamma^{(j)} v_b^{-2},
\end{equation}
or
\begin{equation}
\label{vstar}
v_* = \left[{2 \xi \eta \nu E_0 f_{_{OB}}\over (1-R)}\right]^{1/2}\gg
v_b,
\end{equation}
where $v_*$ is a characteristic velocity associated with the energy
injection by star formation, via SN explosions, into the ISM: for the
parameters adopted in this paper, $v_* \sim 115$~km~s$^{-1}$.  
We now made the following simplifying hypotheses: {\it (i)} the
mass loss induced by isolated SNe is negligible; {\it (ii)} velocity
dispersions are supersonic: $c_{s,eff}^2=c_s^2 + \sigma^2 \simeq
\sigma^2$ (see Fig. 2); 
{\it (iii)} the velocity dispersion has its equilibrium value.
With these assumptions, we can derive an explicit expression for $v_b$
by substituting eq. \ref{eqsigma} into eq. \ref{vb1}:
\begin{equation}
\label{vbeqs}
v_b = 101~L_{38}^{1/3} b_1^{1/6} \phi^{1/4} \varpi_0^{1/3}
M_{g,7}^{(2\alpha-1)/6}~~~~{\rm km~s}^{-1},
\end{equation}

We can now determine the regions in the $(1+\phi)-M_{g,7}$ plane 
where outflows dominate over mass consumption
($v_*>v_b$), along with the blowout ($v_b > v_e$), and blowaway
(eq. \ref{cond}) conditions. In terms of the previous expressions,
these three conditions translate into:

\begin{equation}
\label{bodom}
{\it Blowout} \rightarrow \phi \simlt 5\times 10^{10} L_{38}^4 b_1^{2}
\varpi_0^{4} M_{g,7}^{2(2\alpha-3)} h^{-4/3};
\end{equation}
\begin{equation}
\label{badom}
{\it Blowaway} \rightarrow \phi \simlt 0.27\left[L_{38}^{1/3} b_1^{-5/6}
\varpi_0^{1/3} M_{g,7}^{(\alpha/3-3/2)} h^{-1/3}\right]^{12/19}.
\end{equation}
\begin{equation}
\label{outdom}
{\it Outflow~dominates} \rightarrow \phi \simlt 0.3  L_{38}^{-4/3}
b_1^{-2/3}
\varpi_0^{-4/3} M_{g,7}^{2(1-2\alpha)/3};
\end{equation}

\begin{figure}   
\centerline{\psfig{figure=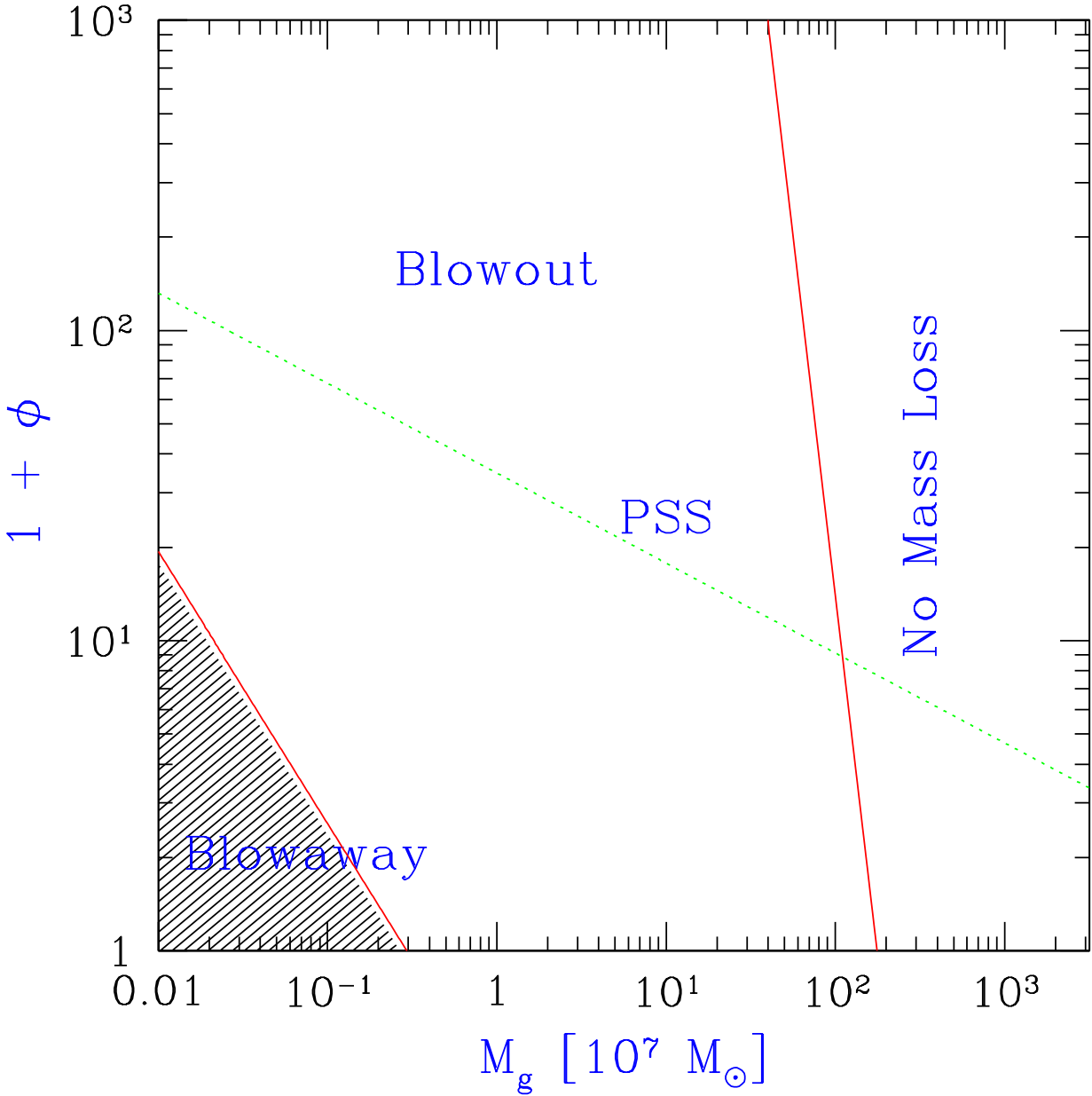}}
\caption{\label{fig3} \footnotesize{Regions in the $\phi -M_g$ plane in which 
different
dynamical phenomena (see text for definitions) may occur, explicitly derived 
in eqs. \ref{outdom}-\ref{badom}.  Also shown is the
locus of points describing the Persic \etal (PSS) $\phi- M_g$ relation
(dotted line).}}
\end{figure}

These relations provide a qualitative understanding of the various
fates of a dwarf galaxy (\ie closed box evolution, blowout, blowaway,
gas consumption) and are graphically displayed in Fig. \ref{fig3}. 
in which we have fixed $L_{38}=1, b_1=1$.
{\it Galaxies with gas mass content larger than
$\approx 10^9\Msun$ do not suffer mass losses, due to their large potential
wells (a result also found in the numerical simulations by MF)}. 
Of course, this does not rule out the possible presence of
outflows with velocities below the escape velocity (fountains) in
which material is temporarily stored in the halo and then returns to
the main body of the galaxy, as discussed in \S~2.  For galaxies with
gas mass lower than this value, outflows cannot be prevented.
If the gas mass is reduced further, and for  $\phi\simlt 20$, a blowaway,
and therefore a complete stripping of the galactic gas, should
occur.  Just as an example, we have plotted the expected value of
$\phi$ as a function of $M_g$, as empirically obtained by PSS (eq. \ref{pss}), 
which should at least give an idea of a likely location of the 
various galaxies in the plane of Fig.
\ref{fig3}. If dwarfs are a one parameter family with respect to these
quantities, they should align along the dotted line, and the
transition from blowout to no mass loss regime should occur at
$M_{g,7}\approx 100$.  

Eq. \ref{outdom} indicates that the predominant mechanism of gas
consumption is provided by the conversion of gas into stars, even
when an outflow takes place. In fact, given the dependence of 
eq. \ref{outdom} on the gas mass, in larger galaxies (\ie $M_g \simgt
10^9 M_\odot$) outflow losses could in principle dominate, but the previous
discussion shows that in such systems blowout - and consequent
mass loss - cannot occur. The reason for this behavior is that 
driving a wind is a relatively inefficient process: the mass
of gas that has to be transformed in stars is always larger than 
the mass of gas ejected by the energy available to drive the outflow
produced by the same stars. 
We conclude that {\it the main source for gas consumption is provided
by star formation}, in agreement with Davies \& Phillips (1988). 

The previous results do not analyze the changes in the gas content of
the galaxy during its evolution; Fig. \ref{fig3} is meant to give a
simple schematic overview of the most important physical processes and
of their approximate domains. The detailed discussion of the fully,
time-dependent results is given in
Sec. 4.

\subsection{A Dark Matter-Modified Schmidt law}

Using relations \ref{nhn}-\ref{n0} we can rewrite eq. \ref{mg2}
explicitly as
\begin{equation}
\label{schm}
{d\over dt}M_g= A {M_g^{3/2} M_T^{1/2} c_{s,eff}^{-1}} - W(t)
\end{equation}
where $A=\sqrt{2}(R-1)G/\tau \varpi^2_*$.  The additional assumption of
a steady state for $\sigma$ allows to eliminate the velocity
dispersion (contained in $c_{s,eff}$)  
from eq. \ref{schm}.  The first term represents the mass
consumption due to star formation, and the second describes the mass
loss in the outflow, if present.  It is interesting to note that the
expression for star formation rate resembles a standard Schmidt power
law, but with the relevant difference that it depends both on the
gaseous {\it and} the total mass. This may have important observational
consequences, as it will become clear from the results presented in
\S~4.  In the absence of DM          ($M_T=M_g$) the star formation
law power index, assuming equilibrium for $\sigma$, would be equal to
$(7+2\alpha)/4=1.92$.

\section{Available Observations}

We are able to perform different tests of our models against data 
available in the literature.  We can compare model 
predictions against a relatively large 
sample of nearby gas-rich dwarf galaxies.
Since our predictions relate to on-going star formation
in galaxies with an ISM, we necessarily select galaxies
which are gas rich. 
The aim is to test as much of the parameter space as possible to
verify the general predictions of the models.
We consider in addition a
uniform sample of low surface brightness, gas-rich galaxies, 
which are 
underluminous systems
for their HI content (van~Zee \etal 1997a,b). 
We finally model in detail the very 
carefully studied nearby galaxy Leo~A,
for which in particular the star formation history is well constrained. 

Combining observations from many different sources has the difficulty
of numerous different definitions. We have tried to make sure that
the data         which we use come from equivalent measures.
We define the 
total mass of a galaxy as the sum of 
the gas mass, the stellar mass, and the DM mass. We thus define
the DM mass to be the 
short-fall between the total mass and the visible mass (stars plus gas)
mass. Measuring the integrated photometric and structural
properties of nearby galaxies is not a trivial task. The size range
goes from arcmins to many degrees, and nearly all of the galaxies
in our sample have low surface brightness which complicates
reliable and  consistent surface brightness measures.
A number of the smallest dwarf irregular
galaxies are dominated by random motions
rather than rotation, which makes disentangling the different
velocity components, and determining the total
mass especially difficult. Sometimes it is unclear if the HI observed
is in virial equilibrium.

\subsection{The ``Skillman'' Nearby Galaxy Sample}

Table~1 compiles the results from a number of relatively well studied 
nearby galaxies.  The selection of these galaxies is such that they 
contain HII regions bright enough to obtain 
spectroscopy for oxygen abundance determinations, \ie  
with published 12$+ \log$ [O/H]. These mostly 
come from Skillman, Kennicutt \& Hodge (1989), hence the name of the
sample, and then a few more recent 
results were also included. The majority of the metallicities in
the Skillman \etal paper did not include errors, so the
values given in Mateo (1998) were used.
Then the literature was searched for the other 
parameters of interest, namely HI mass, total dynamical mass, and 
surface brightness.  It was difficult to find reliable measures of the ISM 
velocity dispersion in these galaxies, and some of the values are of 
dubious accuracy, and with a few exceptions, which are indicated, 
most of the $\sigma_{ISM}$ 
should be treated as upper limits as well.  There were also a number of different 
definitions of surface brightness determinations, which are intrinsically 
difficult measurements for low surface brightness irregular galaxies.  
These, in certain indicated 
cases, should be considered as upper limits.  Often, when 
looked at in detail, a galaxy is found to contain an extremely extended low 
surface brightness halo (e.g.  Minniti \& Zijlstra 1996).  The 
(total) dynamical 
mass of many of the galaxies in this sample 
comes from single-dish observations by Huchtmeier \& Richter 
(1988), using the HI linewidth versus optical diameter relation defined by 
Casertano \& Shostak (1980). 
The footnotes of Table~1 should be carefully 
noted, as they provide an estimate of the true reliability of each measure.

\subsection{The ``Van Zee'' Low Surface Brightness Dwarf Sample}

Another comparison with our model comes from the
detailed self-consistent work of van Zee \etal (1997a, b)
on a sample of galaxies defined by unusually 
large M$_H$/L$_B$ values.
The sample includes isolated HI rich galaxies with extended
HI envelopes located around low optical luminosity spiral
galaxies, and comparison ``normal'' dwarf galaxies.  The normal
dwarfs have similar star formation rates and other global properties to the
low surface brightness sample, but they are less massive.
The gas dynamics and kinematics of most of
these galaxies were then studied in great
detail, as were the young stellar population (via colors and
metallicities of HII regions), and metallicities were also
measured by van Zee.
Thus, this is a good comparison sample for our models even though the
galaxies have slightly higher mass and larger physical scale than
the single cell type galaxies which make up the majority of the
Skillman sample. The values are given in Table 2. There is one galaxy
which overlaps with the Skillman sample (DDO~187).

\subsection{An Individual Case: Leo A}

Focusing even further in on our model predictions we now make a
detailed comparison with the known star formation history (SFH) and
metallicity of a nearby dI galaxy, Leo~A, which is part of the Skillman sample,
and has a even lower surface brightness than typical for the Van Zee sample.
We choose Leo~A because it has one of the most detailed
SFH available for a low surface brightness,
dwarf irregular type galaxy (Tolstoy \etal 1998). 
The SFH is reasonably well
determined from stellar color-magnitude diagrams (CMD) studies. Such studies
(Tolstoy \& Saha 1996, Tolstoy \etal 1998, Tolstoy 1998b) 
are able to accurately trace the SFH using several indicators
derived from theoretical stellar evolutionary tracks. 

\section{Comparisons with Samples}

We first consider 
how the model predictions 
fit in with the global properties
of nearby gas-rich galaxies.

\subsection {Oxygen Abundance}

\begin{figure}   
\centerline{\psfig{figure=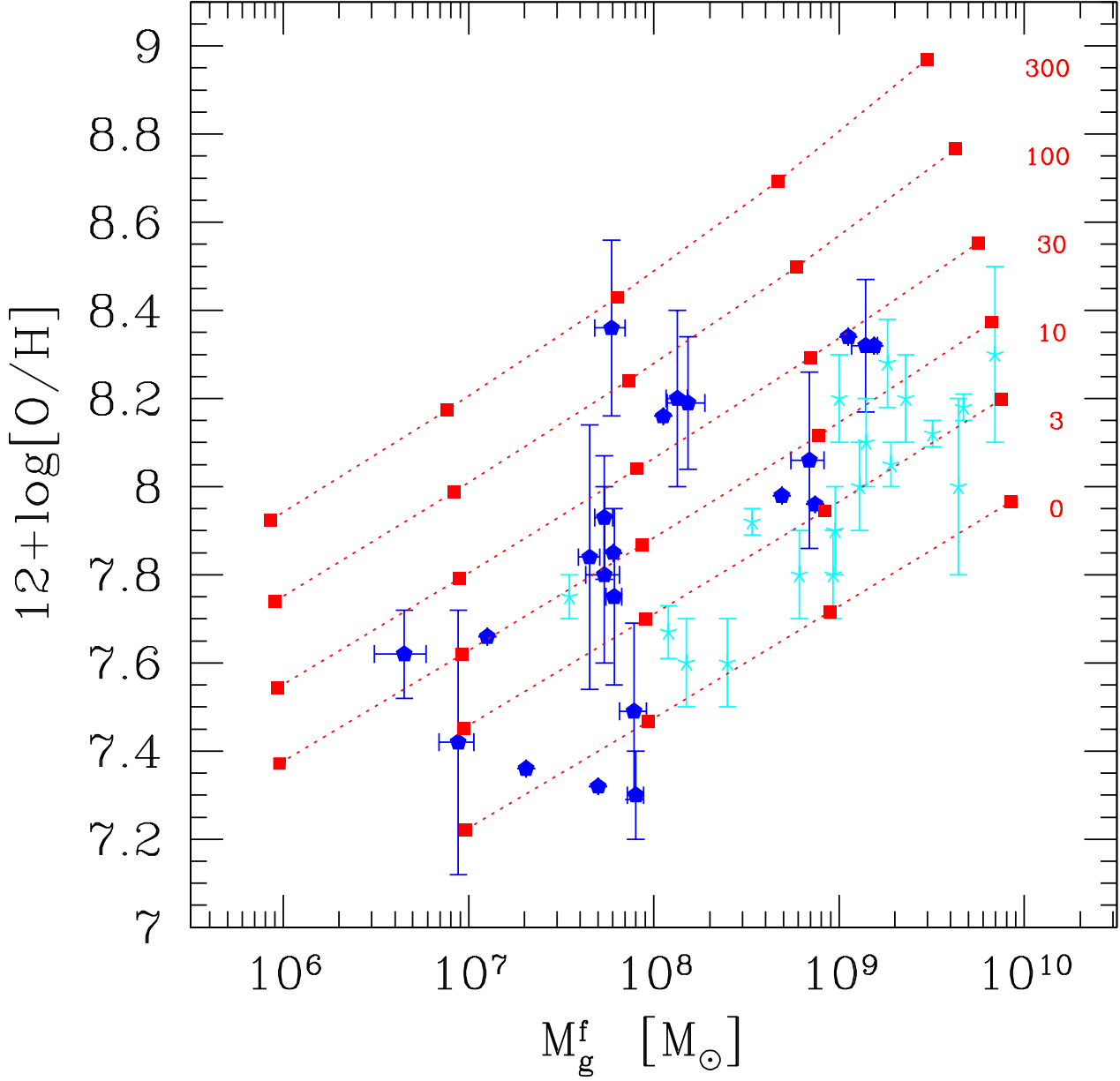}}
\caption{\label{fig4}\footnotesize{ Model results for
the oxygen abundance as a function of the final gas mass $M_g^f$ 
for different values of $\phi=0,3,10,30,100,300$ (squares); 
points show the data from the Skillman (pentagons) and van Zee (stars)
samples.}}
\end{figure}

Fig. \ref{fig4} shows a comparison between the data and the model results for
the oxygen abundance as a function of the final gas mass $M_g^f$ for
different values of $\phi$.  As a general trend, the oxygen abundance
in the samples of dwarf galaxies, listed in Tables~1 and 2, increases
with $M_g^f$: this behavior is correctly reproduced by our
results.  However, for a given $M_g^f$, there is a considerable spread
in the metallicity data. Assuming that galaxy interactions have
not played a significant role in the evolution of any of these
systems, then
this can be understood as an effect of the
different DM content, where {\it 
higher dark-to-visible mass ratios produce higher final values
of $X_o$}.  This is due to the fact that as the DM content is
increased at fixed gas mass, the gas tends to be compressed towards
the central regions by the stronger gravitational potential. As a
consequence of the higher density, the star formation process is
favored and its rate increased. For similar reasons,
adding DM makes outflows, and
hence oxygen mass loss, less efficient. 

\begin{figure}   
\centerline{\psfig{figure=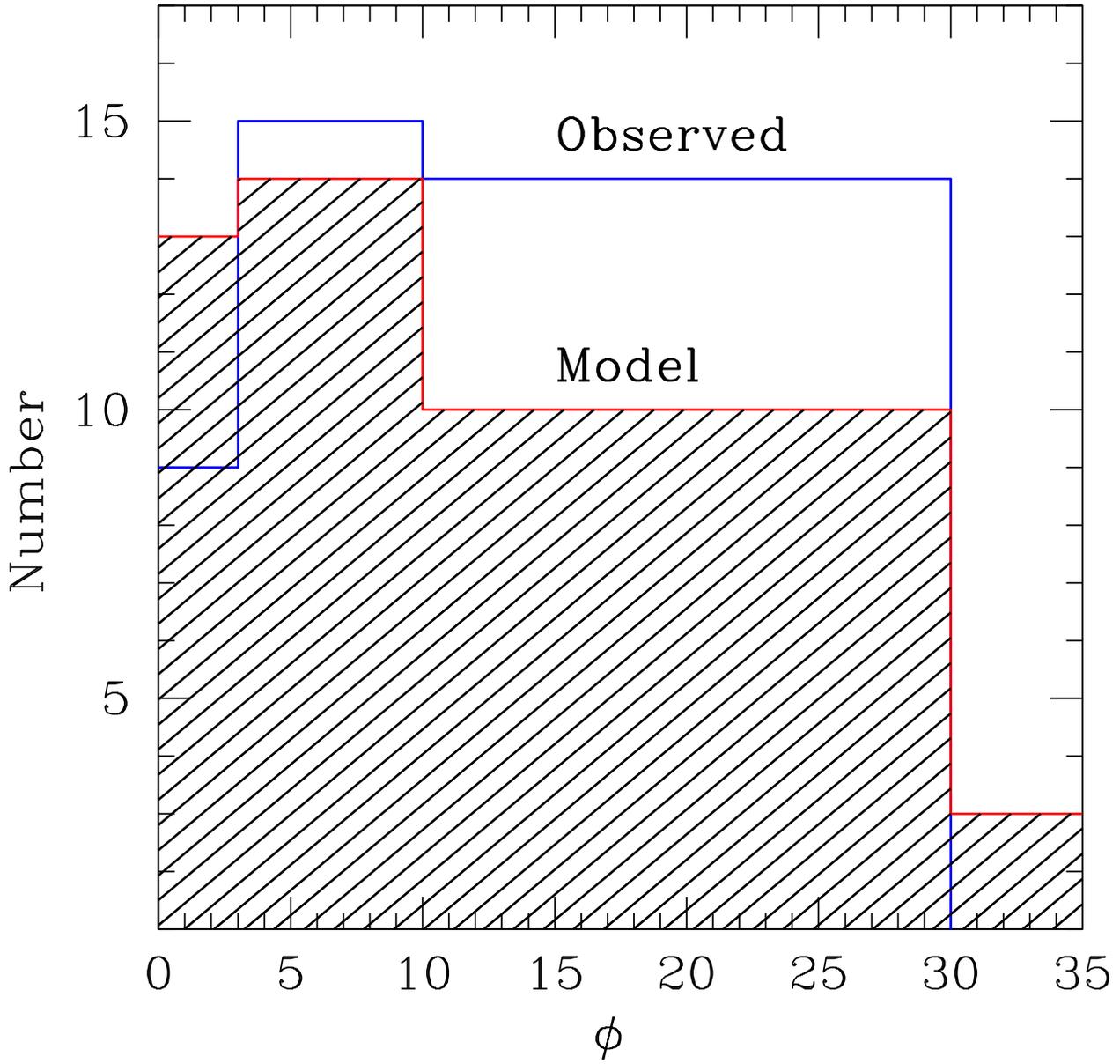}}
\caption{\label{fig5}\footnotesize{ Distribution 
of the observationally inferred dark-to-visible ratio, $\phi$, for the galaxy
compared with the one derived from the model. }}
\end{figure}

From Fig. \ref{fig4} it appears that almost all sample 
galaxies {\it can be inferred to have
dark-to-visible mass ratio $\phi \simeq 0 - 30$} (although about 65\%
of the objects have $\phi \le 10$ as shown by the observed distribution
in Fig. \ref{fig5}), in order to explain 
the observed oxygen abundances. The values of $\phi$ derived    from the model 
are also reported and the two distributions are seen to be in good agreement.

The sample of galaxies have gas masses 
in the range $10^7 - 10^{10} M_\odot$.
The lower limit is consistent with our model limit at which 
less massive objects cannot
survive blowaway, as indicated by the absence of low $\phi$ galaxies
below $10^7 M_\odot$.  For these extremely small objects a
considerable amount of DM is required to prevent blowaway, 
as already noted in Sec. 3.
Thus, a prediction of the model is {\it the existence of a
lower limit for the oxygen abundance} of small mass ($M_g^f\simlt 10^7
M_\odot$) dI: almost independently of $\phi$, $X_o^{min} \approx 0.02$
(\ie $12 + \log(O/H)\approx 7.2$).  
This lower limit is set by the minimum amount of DM
necessary to avoid the blowaway, 
which results in the minimum
metallicity. We        
expect that dI galaxies with $X_o< X_o^{min}$ should not be observed
since they are likely to be disrupted during the initial phase of star 
formation.
It is possible that 
today's dSph/dE galaxies could be the remnants of such a
catastrophic event, consistent with dE galaxies typically being
dominated by
old stars and being extremely metal poor; we will
discuss these issues in more detail in Section 6.

It is interesting to note that, there is a group of objects with
masses around $10^7 M_\odot$ that should be critically
close to fulfill the blowaway condition. 
The fact that these objects are among the
faintest star forming objects known 
is consistent with the destruction of even smaller
galaxies in the past.

\subsection{Mass Outflow Rate}
 
\begin{figure}   
\centerline{\psfig{figure=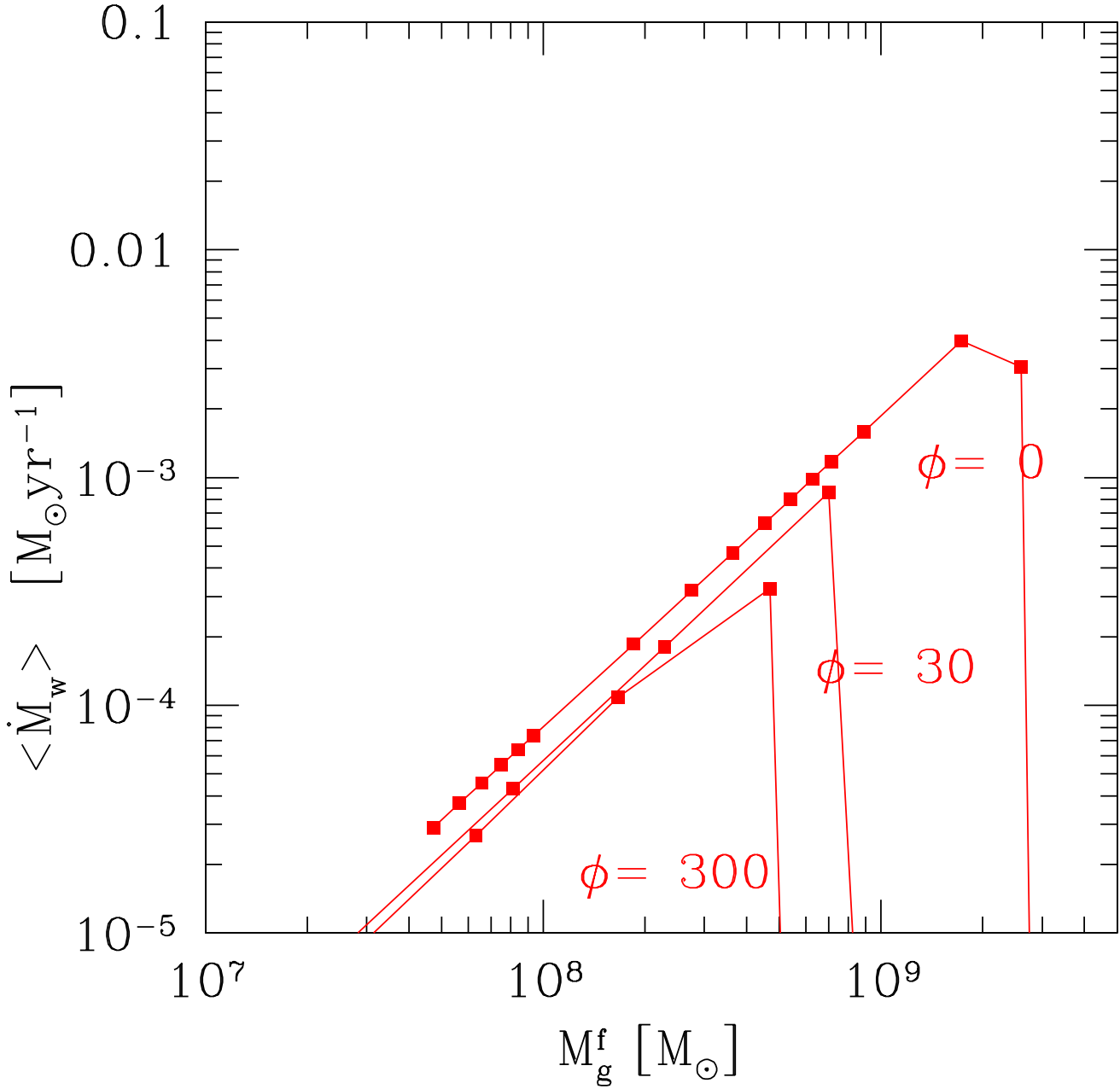}}
\caption{\label{fig6}\footnotesize{Time-averaged mass outflow rate $\langle 
\dot M_w \rangle$ 
as a function of the final gas mass $M_g^f$ for three different values of 
$\phi=0,30,300$.  }}
\end{figure}

Most dwarf galaxies are predicted to evolve through an outflow phase
in which a fraction of the gas is injected in the intergalactic medium
(Wyse \& Silk 1985, Vader 1986, 1987). In general, we find that the
outflow mass loss in a galaxy is a decreasing function of time and that single SN
events are not powerful enough to drive a
significant outflow. Fig. 5 shows the
behavior of the mass outflow rate $\langle \dot M_w \rangle$, averaged
over the entire evolutionary time. 
The {\it mass loss rates are in the 
range $10^{-5}-0.04 ~M_\odot$~yr$^{-1}$}. 
The mass loss is a rather steep
function of gas mass peaking at $M_g^f\approx {\rm few} \times 
10^8 M_\odot$. 
The peak is the product
of two effects: on the one hand the galaxy has to be sufficiently gas
rich to form enough massive stars to power the outflow; on the other
hand, larger masses correspond to higher escape velocities which 
make the outflow increasingly difficult.  
For a given galaxy
gas mass, $\langle \dot M_w \rangle$ is only weakly dependent on the
dark-to-visible mass ratio, as can be appreciated from Fig. \ref{fig6}, which
shows the comparison between the three cases $\phi=0, 30, 300$.
Consistently with the above interpretation, $\langle\dot M_w\rangle$ 
decreases with increasing $\phi$; however, the differences between the
cases $\phi=0$ and
$\phi=300$ are less than a factor $\simlt 10$ in the mass range considered;
at the high mass end the larger $v_e$ quenches
the blowout (see Fig. \ref{fig3}). 
Galaxies of mass ${\rm few} \times 
10^8 M_\odot$ 
are therefore the {\it major pollutants of
the IGM with both mass and heavy elements}.  

\subsection{Gas Velocity Dispersion}
 
\begin{figure}   
\centerline{\psfig{figure=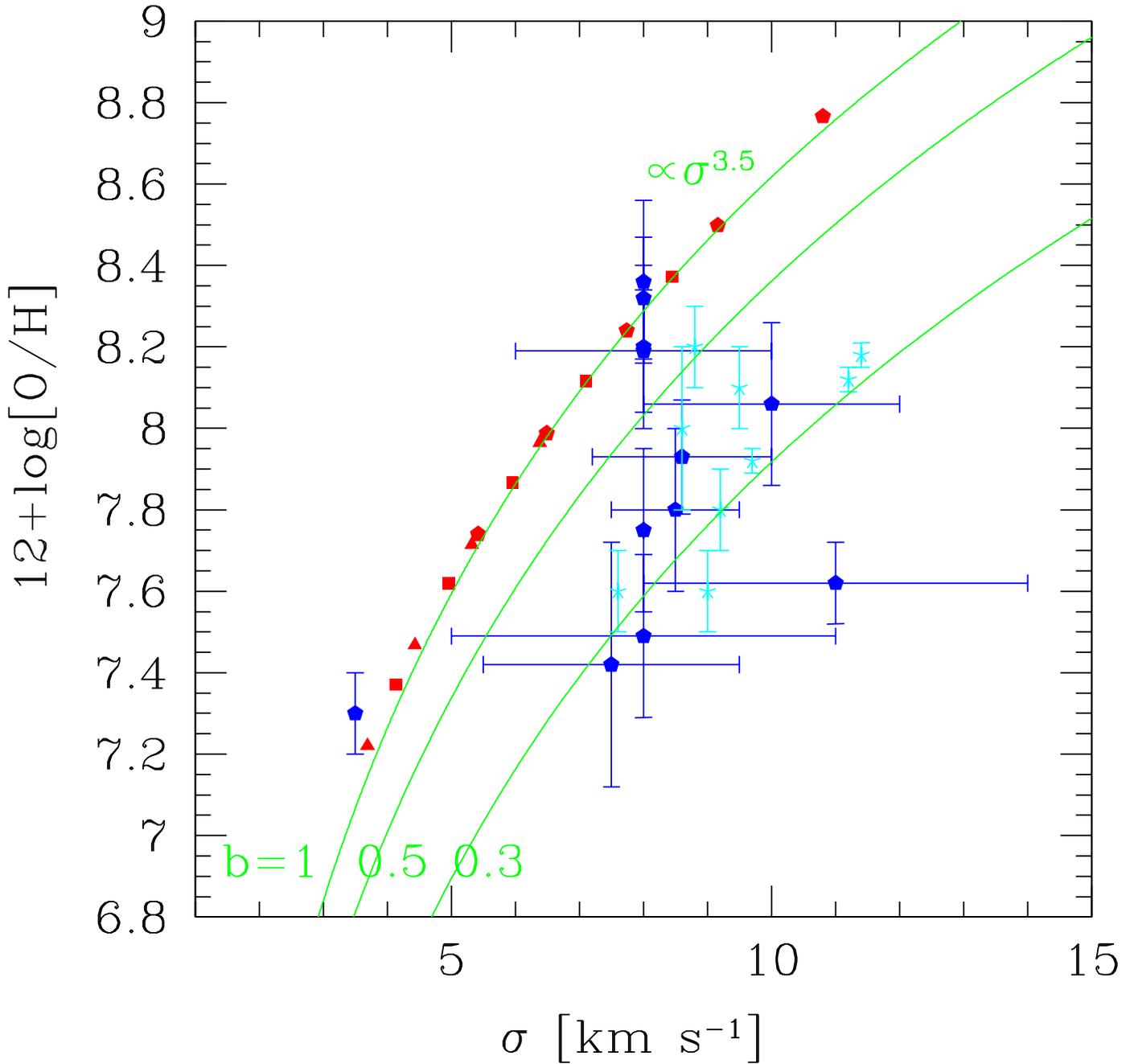}}
\caption{\label{fig7}\footnotesize{ Relation between oxygen abundance and 
gas velocity dispersion for $\phi=0$ (triangles),$\phi=10$ (squares),
$\phi=100$ (pentagons). A fit to the relation is also shown for three
values of the magnetic reduction factor $b_1=1.0,0.5,0.3$ (solid lines).
Data points are from the Skillman (pentagons) and Van Zee (stars) sample.}}
\end{figure}

As SNe are the most important oxygen source,  and they
also regulate the kinetic energy input to the ISM, it is
natural to investigate the relationship
between $\sigma$ and $X_o$ in galaxies.  
Such a relationship, as obtained from our
model, is shown and compared with available data in Fig. \ref{fig7} 
for three different values of $\phi = 0, 10,
100$ which result in different $M_g^f$ (compare with
Fig. \ref{fig4}). This relationship is surprisingly tight, in the
sense that dI galaxies with higher metallicity tend to show higher
velocity dispersions, at least for the uniformly measured
van~Zee sample.  As said above, there are general reasons to
expect such trend, but the interesting aspect is that all galaxies
{\it independently of their DM          content fall on the same
$X_o - \sigma$ curve, which has the rather steep slope 3.5}, as seen
from the fitting curve drawn on top of the points in Fig. \ref{fig7}. Several
independent processes concur to establish this 
relation and their detailed interaction is not easy to
disentangle. Ultimately, however, the most important factor is the
interplay of DM          and SN feedback in regulating both the
chemical and kinetic energy balance of the ISM.  
However, the magnetic field is also an important ingredient as it can  
result in more elastic, and hence less dissipative, cloud collisions.
Inclusion of a magnetic field does not modify  the above relation 
between metallicity
and gas dispersion, but it could shift it to larger values. This 
effect is quantified in Fig. \ref{fig7} where we have drawn 
the analogous  curves for three different values of $b_1$, the 
magnetic reduction factor introduced in eq. \ref{eminus} (this
is the only case in which we explore $b_1$ values different from 1). 
{\it A reduction in the cloud collision energy dissipation rate seems to 
fit better the data, with a preferred value $b_1\approx 0.5$}. This means
that about half of the kinetic energy of the clouds is temporarily
stored into magnetic line tension and it is released again after 
the collision, consistent with the findings
of Miniati \etal (1999).

\subsection{Surface Brightness}

Our models must also be able to
reproduce, at least to first order, the observed
properties of the sample stellar populations. 
Here we start from the surface
brightness - visible (\ie disk gas + stars) mass relation.
At the present stage, we are able to compare our results
with the data only at a basic level since a detailed comparison would
require a self-consistent computation of the photometric evolution of 
the galaxy. However, as a sanity check, it is important to get at
least an approximate estimate of the surface brightness.

\begin{figure}   
\centerline{\psfig{figure=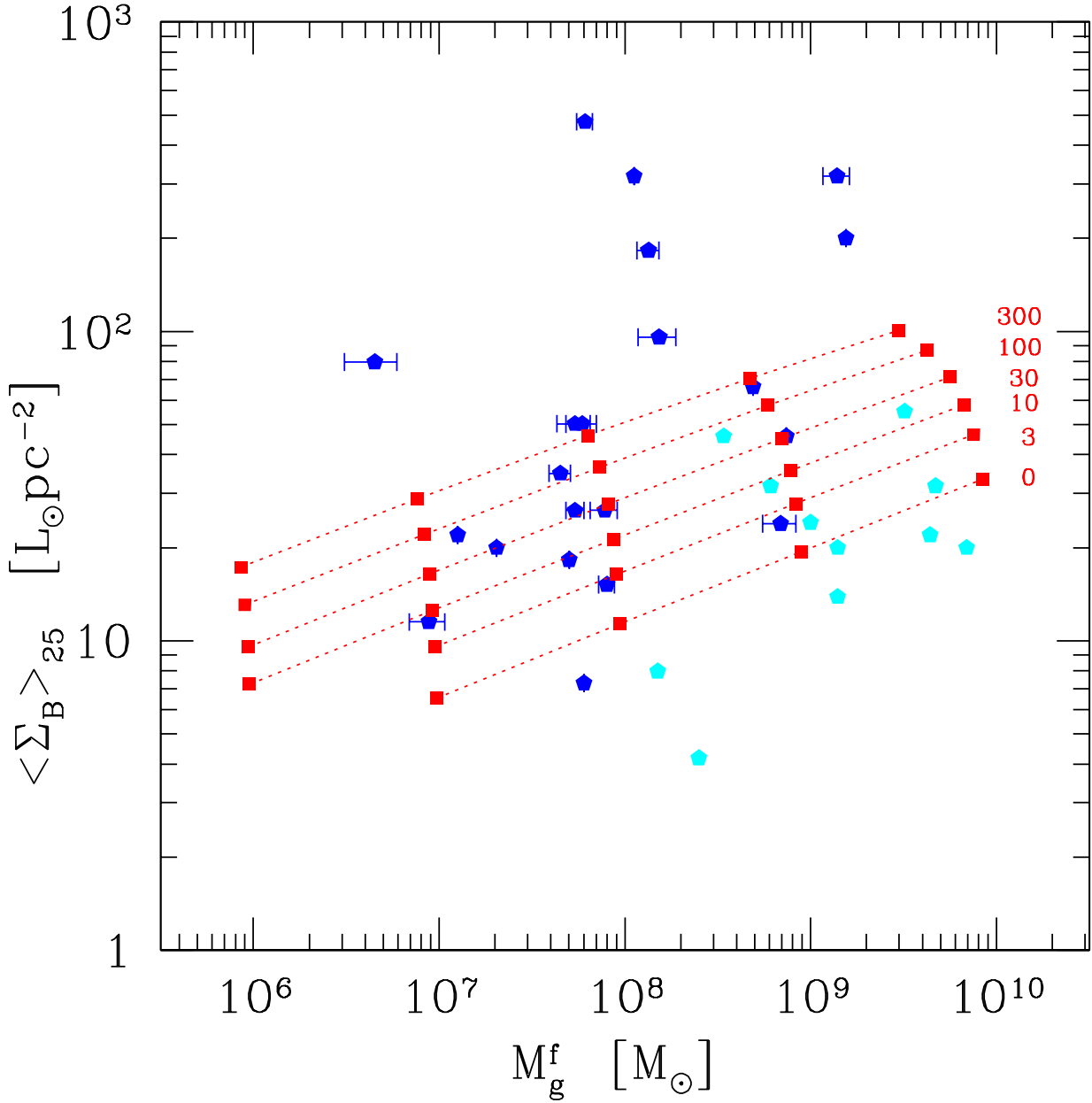}}
\caption{\label{fig8}\footnotesize{ Surface brightness at B=25 
as a function of the final gas mass $M_g^f$ for
different values of $\phi=0,3,10,30,100,300$ (squares); pentagons
show the data from the Skillman sample.}}
\end{figure}

To this aim we use the results obtained by PSS, who have
investigated the structural properties of a sample of 1100 spiral
galaxies using optical and radio rotation curves and relative surface
photometry. According to these authors, the relationship between the
visible mass and the B-band luminosity of a galaxy has the following
expression (solar units):
\begin{equation}
\label{monl}
\log\left(M_v\over L\right)= 0.5 + \left[ 0.35 \log \left(L\over
L_\star\right) - 0.75 \log^2 \left(L\over L_\star\right)\right].
\end{equation}
By applying this formula, and assuming the usual mass-radius relationship
as given by eq. \ref{rc} and a disk mass-to-light ratio of 0.5, we
have derived the surface brightness of the model galaxies. The
comparison between the theoretical and observational data points,
shown in Fig. \ref{fig8} for different values of $\phi$, indicates a
satisfactory agreement within the limitations mentioned above. The
general trend of      increasing surface brightness with gas mass is
successfully reproduced as well as the typical range of values. 
Given the assumptions made, we do not 
emphasize this result any further.
As a final remark, one has to keep in
mind that the extrapolation of the empirical relation eq. \ref{monl} to
the mass range relevant to the least massive dwarfs is at this stage
only very tentative. The upper points in the figure are all ``bursts'',
or BCD type galaxies, which are likely to possess faint underlying
halos which have not been included in the surface brightness measures,
making it an upper limit. As we argue several times in this paper,
these burst episodes affect only mildly the global evolutionary properties
of dwarfs, although the increased luminosity during these periods might
allow their detection at cosmological distances (see Section 6).

\subsection{Star Formation History}

\begin{figure}   
\centerline{\psfig{figure=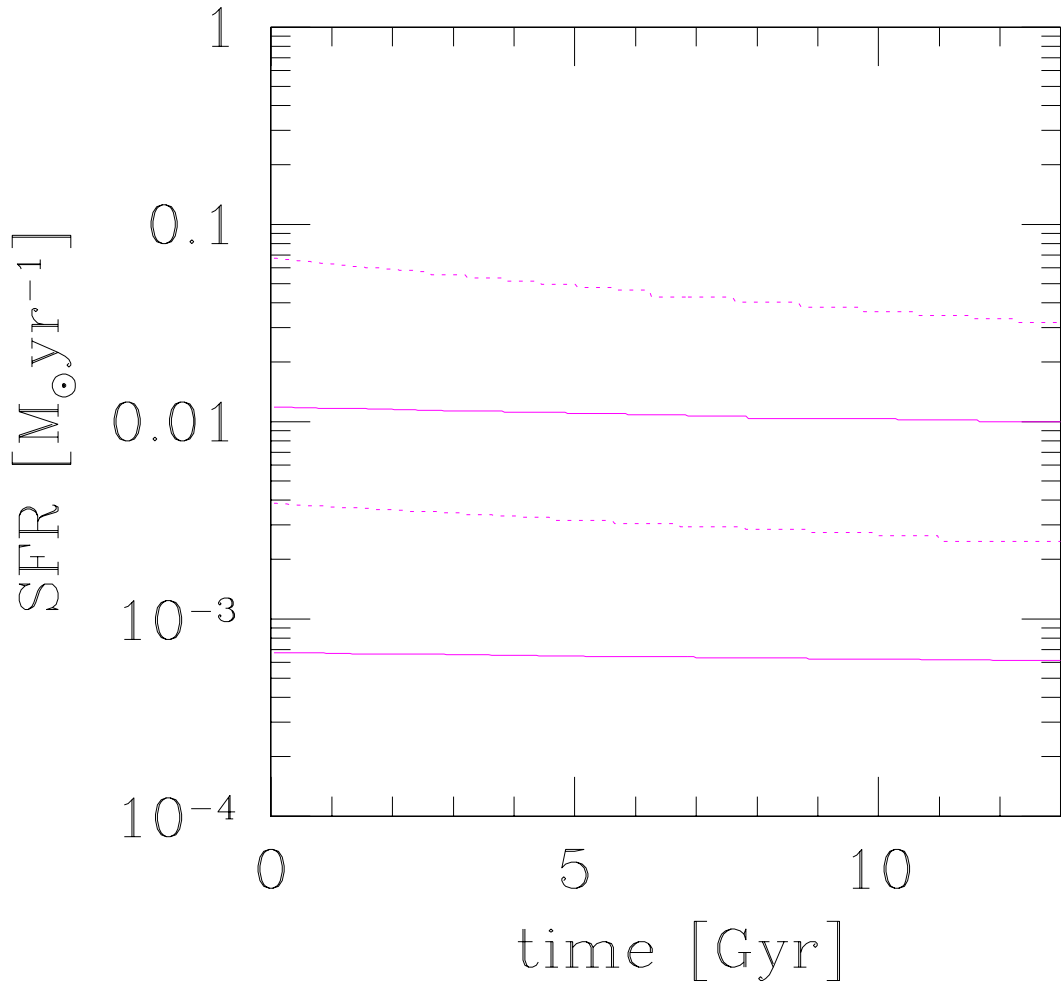}}
\caption{\label{fig9}\footnotesize{ Evolution of the star formation rate for 
four different
galaxies with final gas masses  $M_g^f\approx 10^8 M_\odot$ (solid curves) 
and $M_g^f \approx 10^9 M_\odot$ (dotted); for each mass two values
of $\phi=0$ (lowermost) and $\phi=30$ (uppermost) are shown. }}
\end{figure}

The presence of DM profoundly influences the star
formation history of dwarf galaxies. Its main influence is on
the processes regulating the star formation rate is via the
compression of the cool gas in a deeper gravitational potential, thus 
increasing the mean gas density.  We have computed the star formation
rate for galaxies with different gas masses ($M_g^f \approx 10^8
- 10^9 M_\odot$) and values of $\phi=0, 30$. Our model assumes that
galaxies are evolving in a relatively
undisturbed manner, as we are not considering any external (or
internal) perturbation which could trigger vigorous star forming
episodes. The behavior of the star
formation rate in time for the set of model galaxies above is shown in
Fig. \ref{fig9}. The rates tend to decrease in time as the gas is either
consumed by star formation (predominantly) or lost in the outflow. 
The variation of the gas mass is relatively modest though,
being within a factor 2-3 of the initial value, independently of the
amount of DM. As more DM is added, it increases the star formation
rate. {\it For a $M_g^f \approx 10^9 M_\odot$ the star formation rate is in
the range $2\times10^{-3}-0.07~M_\odot yr^{-1}$, 
depending on the
evolutionary stage and on its dark-to-visible mass ratio}.

\subsection{A Specific Case: Leo A}

\begin{figure}   
\centerline{\psfig{figure=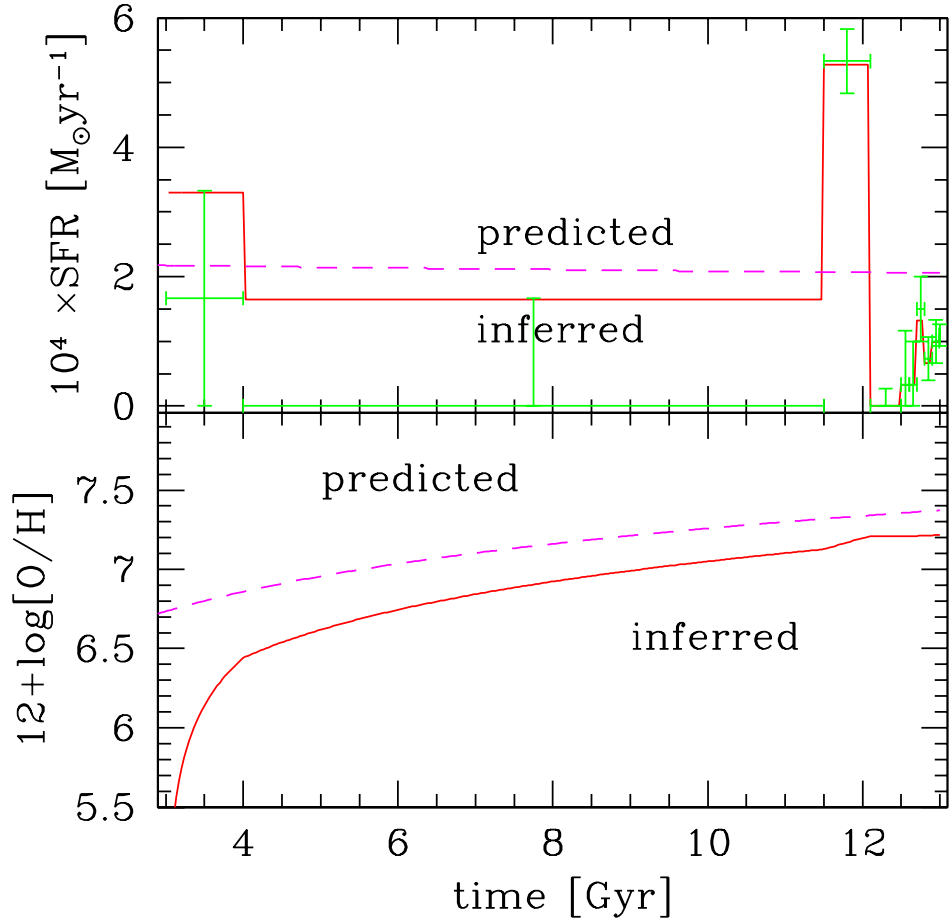}}
\caption{\label{fig10}\footnotesize{ {\it Upper panel}: star formation history 
for the galaxy Leo~A
inferred from CMD diagrams compared with the one predicted from the model;
{\it Lower panel}: evolution of metallicity for the two SFHs. Both give final 
oxygen 
abundances within the errors of the currently observed value 
$12+\log[O/H]=7.3\pm 0.1$. }}
\end{figure}

The SFH obtained by Tolstoy \etal (1998) for Leo~A is shown in the upper panel 
of
Fig. \ref{fig10} (solid curve). Essentially one major star formation
episode ($\approx 5\times 10^{-4} M_\odot$~yr$^{-1}$) appears 
to have occurred about 1.5 Gyr ago lasting 0.6~Gyr, preceded by a long lasting 
low level of star formation similar to that observed today. 
This older star formation activity  is poorly  constrained by current data.
For the long-lasting ($\approx 11$~Gyr)
interval before the burst, only an upper limit of about 
$\approx 1.7\times 10^{-4} 
M_\odot$~yr$^{-1}$ can be set. This is broadly consistent with
variations about a constant (but low) level.

We make two comparisons with our models. We input
the Tolstoy \etal SFH for Leo~A into our model, and predict the
resulting  metallicity. We also compare the SFH and
the final metallicity predicted by our model with the gas mass and
dark-to-visible mass ratio taken from the observations of 
Leo~A, independently  of what is known about the SFH.

For the first case we obtain the metallicity
evolution for Leo~A (see bottom panel of  Fig. \ref{fig10}). In order 
for the final metallicity ($12+log[O/H]
=7.2\pm 0.1$) to agree with the observed value ($12+log[O/H]
=7.3\pm 0.1$, it is necessary to assume that in the latency period 
star formation has proceeded at a rate equal to the upper limit set
by observations ($\approx 5\times 10^{-4} M_\odot$~yr$^{-1}$);
a lower level of star formation would underproduce the 
observed oxygen abundance. 
Alternatively, one could possibly explain the metallicity of the galaxy
by postulating an extremely early burst,
\ie in the first 2 Gyr, which cannot be detected by present observations.
Given these uncertainties, we show only the fit to the data that
is based on available data. The combined metallicity evolution --
CMD information clearly provides useful constraints on the SFH
in galaxies; thus, it will be interesting in the future to check 
if more sensitive observations will allow to decide between the
two above possibilities.

Next, we have run our code with the standard prescriptions, 
\ie without assuming 
the SFH is known {\it a priori},  for the specific case of Leo~A. 
We obtain a metallicity $12+log[O/H] =7.38$. Of course, the derived 
SFH differs from the inferred one (an obvious consequence of the
fact that we have not attempted to allow for significant
variations in star formation rate), but 
it apparently approximates the mean of the one inferred from 
observations and we regard this as a success of the model.   

\section{Remarks and a Possible Scenario}

The model presented in this paper provides a consistent, 
interpretation of the behavior and properties of 
dwarf galaxies. However, it is important to discuss
the uncertainties associated with our analysis.   
There are at least three quantities that we consider poorly 
constrained and which might influence our results: {\it (i)} the HI radius -
mass relation, {\it  (ii)} the strength of the magnetic field 
and {\it (iii)} the oxygen chemical yield. 

\begin{figure}   
\centerline{\psfig{figure=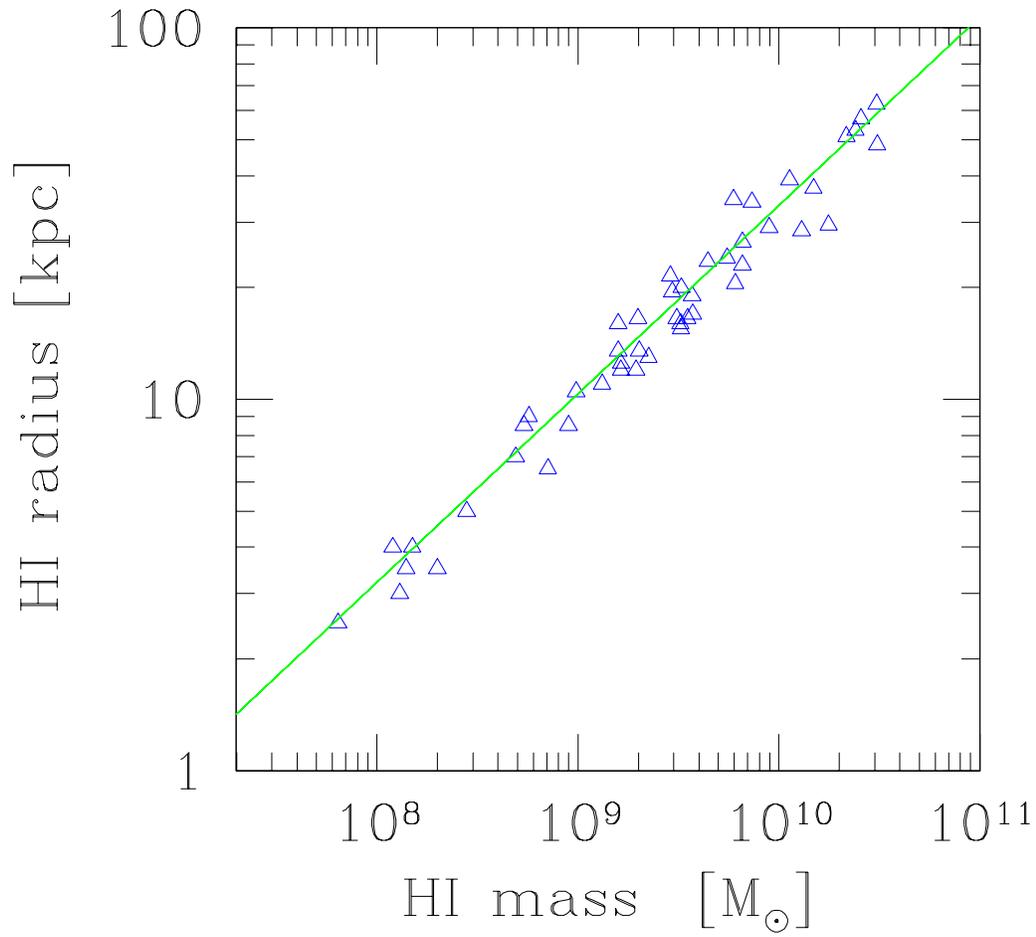}}
\caption{\label{fig11}\footnotesize{ Comparison between the analytical HI 
radius-mass
relation adopted in this work eq. \ref{rc} (solid line) and the data
from Broeils \& van Woerden 1994 (triangles). The line is {\it not}
a fit to the data.}}
\end{figure}

Various authors have found/used different dependences of the
HI radius of dwarfs as a function of their HI mass. For example,
Matteucci \& Chiosi (1983) from a linear fit of a sample containing
45 BCD/dI galaxies derived the dependence $\varpi_* = 0.96 M_{g,7}^
{2/3}$~kpc (using the same factor of 2 as here for the HI-optical 
radius conversion). Although the coefficient $\varpi_0$ is roughly similar,
the curve is sensibly steeper.  A detailed study by Salpeter \& 
Hoffman (1996) of 70 BCDs/dIs yielded a
slope $\alpha \approx 0.5$ for the relation between the geometric mean 
of the outermost HI radius and the optical radius and the HI mass.
The van Zee sample, which is biased towards low surface brightness
objects, has an even flatter slope, $\varpi_* \propto M_{g,7}^
{0.3}$. Clearly these measures are difficult as they depend on the
ability to trace extended HI distributions. 
In spite of these uncertainties, it is reassuring to find an
excellent agreement between our adopted relationship eq. \ref{rc}
and the one derived from the independent sample of 49 spiral galaxies
with extended HI disks by Broeils \& van Woerden (1994). 
The comparison between the data and the analytical relation
eq. \ref{rc} (not a fit!) is shown in Fig. \ref{fig11} and,
although the latter has been derived from a sample of dwarf 
galaxies, it clearly holds for larger objects. Our models are only mildly 
affected by this uncertainty. As a rule of thumb, larger values 
of $\alpha$ with respect to the one adopted here (0.338, see eq. \ref{rc}) 
would produce a flatter metallicity -- mass/luminosity relation; a larger 
$\varpi_0$ would instead shift such a relation towards smaller 
metallicities. Correlation statistics based on larger samples 
are needed to completely clarify this issue. 
  
We have also predicted the existence of a tight relation between gas velocity
and metallicity in dwarfs (it is interesting to note that an analogous 
relationship has been recently found between the stellar velocity dispersion 
and metallicity in four dSphs by Richer \etal 1999). The agreement with 
the data is very good, particularly
if a kinetic energy dissipation during cloud collisions by about a factor of 2
is allowed. This can be easily achieved, under a variety of collision 
conditions
(ranging from adiabatic to radiative cases) if a magnetic field is present.
The elasticity of the collision (defined as the ratio of the final to the 
initial 
kinetic energy of each cloud) is enhanced by the above required factor 2 
for a magnetic pressure equal to about 1/4 of the gas pressure in the ISM 
(Miniati \etal 1999); we recall that this ratio in the Galaxy is around
unity. Our knowledge of the magnetic field strength and configuration in 
dwarfs is still relatively scarce. The most complete survey aimed at the 
determination
of the B strength in dwarf LSBs has been presented by Klein \etal (1992). 
These 
authors, using Effelsberg 100-m observations at 4.75 GHz, report typical 
B strengths $\simlt 2-4 \mu$G for seven dIs in their sample. Although this
conclusion is subject to several caveats due to the assumptions made to derive
it, Klein \etal conclude that these galaxies are probably a scaled down version
of normal spirals as far as their radio-continuum properties are concerned.
Chyzy etal (1997) used the VLA to obtain power and polarization maps of two dIs
(NGC4449 and IC10) and they were able to probe the existence of large scale 
magnetic field patterns, although of uncertain strength. Theoretically, one 
would 
expect magnetic fields of lower strength in dwarfs galaxies than in spirals 
as dynamo amplification ought to be less efficient due to the lower rotation 
speeds.
Thus, a factor of 4 lower magnetic pressure in dwarfs, as required by the best 
fit to 
the velocity dispersion-metallicity data, is certainly reasonable and 
suggested 
by observations, although additional intense experimental efforts will be 
necessary in the 
future.

The oxygen yield from massive stars must still be considered as 
uncertain in spite of a number of recent detailed studies (Arnett 1978,
Woosley \& Weaver 1986, Thielemann \etal 1992, Maeder 1992). A nice 
comparison among the predictions of different theoretical models 
concerning this quantity can be found in Wang \& Silk 1993 
(see Fig. 1 of their paper). 
All models are in satisfactory agreement on the yields from 
stars with masses $\simlt 30 M_\odot$, but they differ by up to 
a factor $\approx 10$ for stars as massive as 80 $M_\odot$,
depending on the metallicity. This difference, as already pointed out
above, is due to the poorly known mass loss rate dependence on
metallicity for massive stars and to the possible formation of
a black hole. Fortunately, this variability range is much narrower
when integration over a reasonable IMF is performed, due to the
rapid decrease of the number of very massive stars.
Nevertheless, the adopted Maeder yields 
depend on metallicity and a comparison of our results with 
the ones obtained using the older Arnett prescriptions still 
produces a discrepancy of about 30\% in the final oxygen 
abundance of the galaxy. This should be kept in mind when 
interpreting the comparison of our models with the data.

In addition to the uncertainties outlined, either caused by a 
experimental difficulties and/or by a limited theoretical 
understanding of various processes, there are obvious 
limitations and shortcomings of the present model. 
Maybe the most evident one is the lack of an attempt 
to model the possible occurrence of enhanced star formation
periods (\ie starburst) during the evolution of the dI galaxies
studied here. The motivation for this choice is twofold.
Theoretically, the origin and characteristics of starbursts phenomena  are
not well understood: a large number of studies have tried to model these 
phenomena with limited success; it is beyond the frame of this work 
to review them properly.  Observationally, there is no clear evidence 
that the majority of dIs go through starburst phases and
that starbursts produce relevant evolutionary 
effects in addition to the ones produced by the underlying more 
quiescent star formation activity. The issue is nicely reviewed 
by Meurer (1999) who pointed out that the observed central surface brightness
of bursting dIs is well below the one measured in BCDs: this indicates that 
dI starbursts are not even intense enough to be recognized as starbursts in 
many cases.
In addition, a considerable fraction of BCDs (about 20\%) shows no 
structural 
evidence (\ie they have nearly exponential $\mu$ and flat color profiles) for
starbursts. Finally, as we have pointed out repeatedly here and in MF,          
the fractional ISM loss is modest even in starbursting prototypes
as NGG1705 (Meurer \etal 1992). These remarks suggest that neglecting the role 
of
starbursts is not seriously affecting the results presented here for dwarf 
galaxies, 
as also witnessed by the correct reproduction of the metallicity range 
observed for
these objects.

Outflows rates of the order predicted here ($10^{-4}-10^{-3} 
M_\odot$~yr$^{-1}$)
may be traced and studied by X-ray and optical emission-line studies. 
Detailed numerical simulations by MF                        have shown 
that gaseous halos with sizes of tens of kpc are
produced, with regions of high X-ray emissivity close to the galactic
disk.  Relatively cool, dense filaments also occur near the galaxy,
well within the external shock, due to shell fragmentation, which are 
detectable by high spatial resolution experiments.
In fact, the bulk of the observed X-ray emission may come from conductive
interfaces associated with these structures, in which the emission is 
particularly
enhanced by nonequilibrium effects.
Detection via absorption line studies towards background objects such
as QSOs has been also demonstrated (Bowen \etal 1997). Of course this 
type of experiment is limited by the low
column densities in a large, rarefied halo, as well as the
availability of suitably located, sufficiently bright, 
background objects, which are scarce 
for small dwarf galaxies. If these difficulties can be overcome,
this technique might provide unique information about the physical conditions
prevailing in the outflows, and more generally, in dwarf halos.

\subsection{A possible scenario}

The results obtained in this study, together with previous
theoretical and observational work, allow
us to sketch a global evolutionary scenario for dwarf 
galaxies. 

\begin{figure}   
\centerline{\psfig{figure=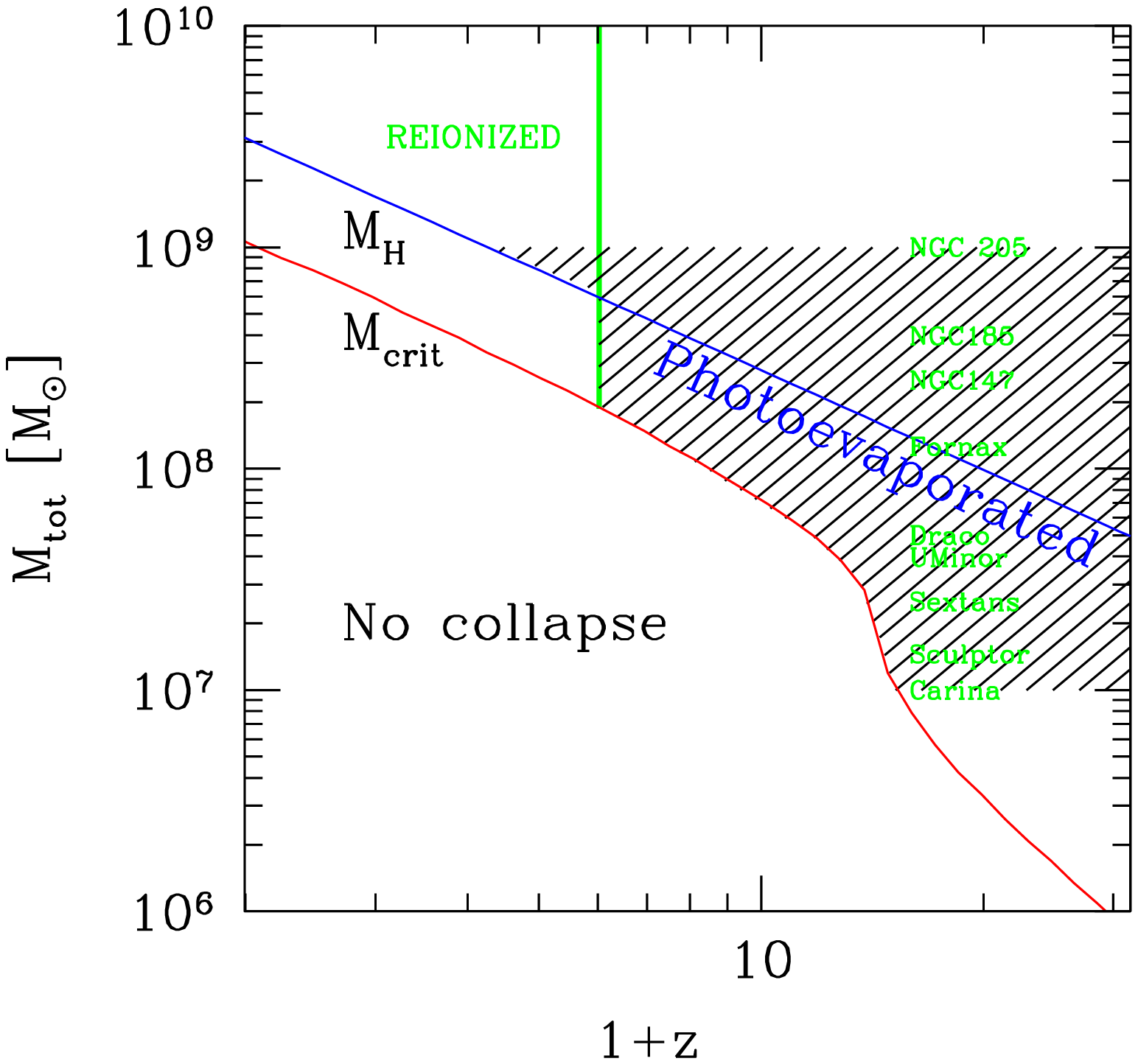}}
\caption{\label{fig12}\footnotesize{ Redshift evolution of the critical mass 
for collapse,
$M_{crit}$, and for photoevaporation, $M_H$. The dashed region corresponds to
the mass-formation redshift space of objects with total masses in the range of 
LG dSphs, whose names are listed. Objects with 
$M<M_H$ can form but they are photoevaporated at reionization (vertical line) 
or blown away; objects with $M>M_H$ can form even after reionization
and survive photoevaporation.}}
\end{figure}

According to standard hierarchical models of structure formation, 
the star formation activity in the universe started with 
the formation of the so-called Pop III objects (Couchman \& Rees 1986,
Ciardi \& Ferrara 1997, Haiman \etal 1997, Tegmark \etal 1997, Ferrara
1998, Ciardi, Ferrara \& Abel 1999) at redshift $z\approx 30$. These are small 
(total mass $M \approx 10^6 M_\odot$ or baryonic mass $M_b \approx 10^5 
M_\odot$) 
objects characterized by virial temperatures, $T_{vir}$,
below the critical one ($T_H \approx 10^4$~K) at which the cooling 
necessary for the collapse cannot be provided by hydrogen lines (mostly
Ly$\alpha$ line radiation), and
molecular hydrogen is then, in a primordial plasma, the only efficient 
cooling agent. Translating the critical virial temperature into a total mass
we obtain $M_H(z) = 4.4 \times 10^9 (1+z)^{-3/2} h^{-1} M_\odot$,
where $z$ is the virialization redshift.
In order for collapse to take place  
it is required that the cooling time (whatever species is providing it) 
is shorter than the Hubble time at the epoch of virialization. 
This defines a second critical mass, $M_{crit}(z)<M_H$, which has been calculated
by Tegmark \etal (1997). As a results only objects more massive than $M_{crit}$
can indeed collapse and form luminous objects.
The situation is somewhat complicated by feedback effects,
involving the photodissociation of molecular hydrogen in Pop IIIs due to 
radiation, possibly created by neighbor sources. This prevents a fraction
of objects, in spite of their mass being larger than $M_{crit}$, from 
collapsing.
The strength of the feedback depends very much on the intensity of the
radiation field and this involves detailed modelling out of the scope of this
paper (these calculations are presented in Ciardi \etal 1999); neglecting this
complication is not crucial for the argument we make here. 
The redshift evolution of both $M_H$ and $M_{crit}$ are shown 
in Fig. \ref{fig12}.

We can now compare these two masses with the ones derived for local dwarfs.
The total masses of the best studied dSphs (Mateo \etal 1993 present the
cases for 9 Local Group dSphs) range from $M_{tot}=1.1\times 10^7
M_\odot$ (Carina) up to $M_{tot}=1.1\times 10^9 M_\odot$ (NGC205).
According to the previous definition $T_{vir}<T_H$, some of them can be classified
as Pop IIIs, whereas the more massive ones probably belong to the class of 
objects able to cool
via Ly$\alpha$. Clearly, as $M_H$ is redshift dependent, the
classification depends on the formation epoch. 
The majority of Pop III objects can only             live 
very short lifes due to their fragility, and the dramatic feedback
effects which can be 
induced by only a handful of SNe; their fate is inevitably
a sudden death after the first star formation episode (which in
most cases only involved the production of $< 1000 M_\odot$ of stars)
and their leftover is a tiny naked stellar cluster without
gas (but with DM), which could possibly be incorporated in larger structures. 
The ones which escape merging could still be seen in the nearby universe,
although their expected extremely low luminosity should make their
detection very challenging.

Is the absence of gas in these objects produced by blowaway in the early 
stages of their evolution ?
The possibility that dEs are remnants of blown away PopIII objects has been 
proposed 
by Miralda-Escud\'{e} \& Rees (1997). The results found here            
might suggest that this interpretation faces a mass discrepancy problem: 
we found that blowaway
can occur only for total masses below $M_c \approx 5\times 10^6 M_\odot$, 
which is
several times smaller than the values derived by Mateo \etal for dSphs.
One has to keep in mind that this value has been derived by assuming galaxy 
properties derived
from observations of present day dwarfs and it implies that gaseous dwarfs,
like dIs, with larger masses                 cannot transform into dEs {\it
in the future}. However,  
the conditions at the formation, as the size, gas density, dark-to-visible
mass ratio, of dwarfs' progenitors were probably very different.
In particular, the critical mass for blowaway is extremely sensitive
($\propto \sigma^{10}$) to the level of turbulent motions in the ISM;
finally,     these quantities depend on the formation redshift and cosmological
model.
This discrepancy might be solved by putting the blowaway 
problem in a cosmological context. In fact,  
specific predictions for the Cold Dark Matter 
model with $\Omega_1, h=0.5, \sigma_8=0.65$, have been made by Ciardi \etal 
(1999) 
who found that the critical mass for blowaway can be increased up to 
approximately
1-2 times $M_H$.

Although some uncertainty remains on the correct value of $M_c$ as a result
of the poorly known formation properties of galaxies, it is clear that
blowaway is a very likely event in these small objects.
This explanation is also favored by current galaxy formation 
scenarios. For example, Klypin \etal (1999) notice that most
popular cosmological models would overpredict the number of low mass satellites
in halos of galaxies like our own by a factor of about 4 with respect to 
what is observed.  Theory can be brought in agreement with observations  if 
it is assumed 
that these missing satellites are dark, \ie invisible due to their faintness  
as a result of gas ejection and quenching of star formation. 
Similar conclusions are reached also by Ciardi \etal (1999) who
find that a consistent number of dark objects should be created as a byproduct
of early galaxy formation. 
Additional support for the occurrence of blowaway has been brought
by an interesting approach recently developed by 
Hirashita \etal (1998). These authors noted that a discontinuity is seen in
the dependence of the virial mass/luminosity ratio (analogous to our $\phi$
parameter) on the virial mass in a sample of local dSph at $M \approx 10^8 
M_\odot$.
They interpret this discontinuity as produced by blowaway, as low mass objects
having lost most of their baryons would be characterized by high values
of $M_{vir}/L$.               

In addition to blowaway, other processes may be able to sweep away the galaxy 
ISM.
Photoionization, either produced by external (Barkana \& Loeb 1999)
or internal (Lin \& Murray 1992, Norman \& Spaans 1997, Spaans \& Norman 1998) 
sources, might have
an important impact on the evolution of dwarfs. If a UV background (UVB), 
produced
by QSOs and galaxies, is present it might even suppress the formation of 
dwarfs (Babul \& Rees 1992, Efstathiou 1992, Thoul \& Weinberg 1996, Nagashima
\etal 1998).  The common physical process on which these two ideas are based 
is that as the gas is photoionized,
it is also heated to a temperature that exceeds the virial temperature of the
halo itself and cannot be confined by the galaxy gravitational potential. 
The photoionization temperature of a moderately dense gas is, for a large range
of properties of the radiation field and medium density, $T\approx 10^4$~K.
Thus, the critical mass relative to  photoionization effects essentially 
coincides with  $M_H$: of course, this is not surprising as both temperatures 
are regulated by Ly$\alpha$ line cooling, which is an excellent thermostat. 

Hence, dSphs must form before a consistent UVB is present in order to be able 
to accrete the gas, \ie before reionization; subsequently they can lose it 
when the UVB intensity has increased. In Fig. \ref{fig12} 
the obliquely
dashed region shows the mass-redshift parameter space were the formation of 
luminous objects regulated by the processes described above is allowed. The 
mass bounds of the shaded region are determined by the upper
and lower mass limits of the 9 dSphs of the Mateo \etal (1993) sample, whose
total mass values are superimposed (clearly, we do 
not know the virialization redshift of these objects).
The first conclusion
that we can draw from the Figure is that objects like Carina, Sculptor, 
Sextans, UMinor, and Draco are indeed Pop III objects which must have 
formed before $z\approx
15$ due to their low masses. Then at reionization, for which we have 
assumed the
conservative lower redshift limit $z_i=5$, they must have lost their gas 
either by
photoevaporation or blowaway. An exception is represented by Sculptor
in which a limited amount of HI ($M_{HI}\simgt 3\times 10^4 M_\odot$)
has been detected (Carignan \etal 1998). This gas is located at the edge
of the galaxy optical image and its association with the galaxy
is not completely clear. This gas could have been either part of a
self-shielded and hence neutral intergalactic clump, or accreted recently thanks 
to the drop in the UV background, or IGM pressure confined ejected gas. 
Larger objects (Fornax, NGC147)
could have formed at redshifts much closer that the reionization one and 
photoevaporated shortly after. NGC 185 and NGC 205 are instead consistently 
found to have some gas (Young \& Lo 1997).
This interpretation still has to face the question why dIs with total 
masses similar
to dSphs (for a comparison see Tab. 1 and 2) yet are gas rich. 
If they have lost
their gas in the past, 
they must have reaccreted material relatively recently when the UVB intensity
dropped. Environmental effects might play a role here, as we must require that
dIs populate regions where gas accretion is more easily achieved  
than in those where dSphs of the same mass are found. 

Whatever process (blowaway/photoionization) is setting the mass
value $M_c$ for gas ejection, might this fact be responsible for the 
separation between late
type and early type dwarfs ? In other words, is it possible that 
these two galaxy types had common progenitors in the past but they
evolved along different evolutionary paths determined by their
mass and hence by their ability to retain their 
gas ? This is certainly an intriguing possibility which would 
solve the problems caused by a dI $\rightarrow$ dE transition 
occurring later on. If, as in the latter case, the transition has 
to be catalyzed by an intense stellar activity expelling the gas, 
then we should look at BCDs as objects on the point of 
being transformed into dEs. There are several, apparently insurmountable,
difficulties associated with this interpretation, which can be 
briefly summarized as follows: although BCDs have clear 
similarities with dIs, however they have {\it (i)} higher surface 
brightness, {\it (ii)} larger radii with respect to dIs (Patterson \&
Thuan 1996); in addition, van Zee \etal (1997a,b) point out that 
BCDs have higher central mass concentration in both gas and stellar 
content. Even the second step of the dI $\rightarrow$ BCD $\rightarrow$
dE chain is problematic: the rotation curves of dIs are noticeably
different from the ones of both dEs and BCDs, their kinematics 
being dominated by chaotic motions (Ferguson \& Bingeli 1994).
The alternative scenario proposed here, in which dI $\rightarrow$ dE 
transitions do not occur but these objects, starting from
the same progenitors, have followed 
different evolutionary tracks, does not suffer from
the above problems. In our view, BCDs would be simply represent 
a "normal" tail of the distribution of dwarfs massive enough to escape both
blowaway/photoevaporation, in addition characterized by higher central
mass density which favors the onset of occasional starbursts.

The consistency of our results for Leo~A between effectively constant
star formation rate over time and ``bursty'' behaviour shows that
the properties of dI and BCD galaxies can be understood as fluctuations
about a mean which is broadly constant/slowly decreasing, value
with time. Bursts of star formation 
can happen easily in these small galaxies, but they
do not last long enough to dramatically overwhelm the long term
effects of low-level, but constant star formation, which is likely
to represent the dominant mode of star 
formation (see also Tolstoy 1996,
Cole \etal 1999, Gallagher \etal 1998). Intense bursts of star
formation  can make these small galaxies visible at cosmological 
distances, and their short duration can make the number density
consistent with that of the faint blue galaxy population seen at 
intermediate redshifts (e.g. Tolstoy 1998a). 
The great diversity of properties of nearby dwarf galaxies
(e.g. Mateo 1998, Toltoy 1998b) indicates that star formation 
rates can and do fluctuate noticeably on short timescales. 
It is also clear, however, that there must be regulatory 
feedback processes, presumably like those presented here, 
which effectively average out these fluctuations over periods of several
hundred million years. Although this is seemingly a long time on the
timescale of an individual HII region, it is a more reasonable
timescale for fluctations in the ISM and it is a negligible 
fraction of the lifetime of a galaxy.

\section{Summary of the results}
We have studied the role of stellar feedback produced by 
massive star formation and DM          on the evolution of
dwarf galaxies. The main results are the following:

$\bullet$ The entire gas content of a dwarf galaxy can be
blown away if its total mass is $\simlt 5\times 10^6 M_\odot$; outflows 
occur in dwarfs with gas masses up to $\simeq 10^9 M_\odot$. Larger 
galaxies are not predicted to have a net mass loss, although
a local circulation of gas via  a fountain-type flow is 
possible. These conclusions closely agree with the 
results of MF.                           

$\bullet$ Even in the presence of an outflow, the predominant 
mechanism of gas consumption is the conversion of gas into
stars.  We conclude that outflows are affecting the properties
of dwarfs only weakly.

$\bullet$ For a given visible galactic mass, the DM          content 
correlates with the gas metallicity; from the available data we
conclude that metallicities are consistent with a dark-to-visible
mass ratio in dIs $\phi \simeq 0-30$. We predict the existence
of a lower limit for the oxygen abundance in dIs 
12+log(O/H) $\approx 7.2$.

$\bullet$ Outflow rates from galaxies, when they occur, are 
in the range $10^{-5}-0.04 ~M_\odot$~yr$^{-1}$; 
The mass loss is a rather steep
function of gas mass peaking at $M_g^f\approx {\rm few} \times 
10^8 M_\odot$; these objects are therefore the major pollutants 
of the IGM.

$\bullet$ The HI velocity dispersion correlates with
metallicity: independently of the DM          content
we predict that $\sigma \propto Z^{3.5}$ . This relationship is
found to agree nicely with the available data, particularly
if about 1/4  of the ISM pressure is contributed by magnetic fields.

$\bullet$ We predict starf formation rates in the range $2\times 10^{-3}-
0.07~M_\odot$~yr$^{-1}$ for a galaxy with present day gas mass $M_g^f \approx
10^9 M_\odot$, depending on its DM          content. 

$\bullet$ For the specific case of the nearby dI Leo A, we forced the 
SFH of the galaxy to be the one experimentally derived from CMD studies; 
with this prescription our model correctly reproduces the observed 
metallicity of the galaxy. If the SFH is instead modelled as in the 
rest of the paper, the final metallicity
is equally well in agreement with the data.

$\bullet$ Based on our results, we discuss a scenario in which late type and
early type dwarfs had common progenitors in the past, but the different
total (mostly dark) mass produced a different evolution, governed by the
combined effects of blowaway of the gas and photoionization. We consider 
dI $\rightarrow$ dE transitions (possibly through a BCD phase) occurring at 
present cosmic times as unlikely. 

\acknowledgments
We would like to acknowledge useful discussions with
B. Benjamin, D. Bowen, J. Dickey, D. Garnett, M.-M. Mac~Low, M. Mateo, 
T. Thuan, and L. van Zee.

\newpage
\small
\begin{table}
\begin{center}
\title{Table 1: Sample Dwarf Galaxy Properties}
\vskip 0.5cm
\begin{tabular}{lccrccl}
\hline
\hline
Galaxy & Z\tablenotemark{a} & $M_{HI}\tablenotemark{b}$ & 
$M_{tot}$\tablenotemark{b} & $\varpi_*$\tablenotemark{b} & 
$\Sigma_B$\tablenotemark{c} & $\sigma_{ISM}$\tablenotemark{c}\\ 
& 12+log(O/H) & $10^{7}M_\odot$ & $10^{7}M_\odot$ & kpc
& L$_\odot pc^{-2}$ & km s$^{-1}$\\
\hline
\hline
NGC2366 & 7.96         & 74.13       & 180 & 7.73  & 45.9 &-\\ 
NGC1569 & 8.16         & 11.22       & 50  & 3.94  & 317.4&-\\ 
NGC4214 & 8.34         & 112.20      & 320 & 12.47 & 0.87 &-\\ 
NGC4449 & 8.32         & 154.88      & 620 & 8.06  & 200.3&-\\
UGC4483 & 7.32         & 5.01        & 10  & 1.21  & 18.3 &-\\ 
\hline
DDO167  & 7.66         & 1.26        & 7   & 1.92  & 22.  &-\\ 
DDO187  & 7.36         & 2.04        & 10  & 1.76  & 20.  &-\\ 
DDO47   & 7.85         & 6.03        & 140 & 5.20  & 7.3  &-\\ 
SMC$^\dagger$    & 7.98         & 48.98       & 50  & 4.93  & 66.3 &-\\ 
\hline
SagDIG$^\dagger$ & 7.42 $\pm$ 0.3 & 0.88$\pm$ 0.19& 0.96& 0.83 &11.52$^e$&7.5 
$\pm$ 2\\ 
GR 8$^\dagger$   & 7.62 $\pm$ 0.1 & 0.45$\pm$ 0.14& 6.0 & 0.96 &79.73$^e$&11 
$\pm$ 3\\ 
Leo A$^\dagger$  & 7.30 $\pm$ 0.1 & 8.0 $\pm$ 0.8 & 6.0 & 0.98 & 15.2 
&9.3$^\ast$/3.5$^\ast$\\ 
WLM$^\dagger$    & 7.75 $\pm$ 0.2 & 6.1 $\pm$ 0.6 & 40  & 2.75 &476.0$^e$&8\\ 
IC5152$^\dagger$ & 8.36 $\pm$ 0.2 & 5.9 $\pm$ 1.1 & 30  & 1.95 & 50.3 &8\\ 
IC1613$^\dagger$ & 7.8  $\pm$ 0.2 & 5.4 $\pm$ 1.1 & 79.5& 2.94 
&50.31$^e$&8.5$^\ast$ $\pm$ 1\\ 
Sex A$^\dagger$  & 7.49 $\pm$ 0.2 & 7.8 $\pm$ 1.3 & 39.5& 3.12 &26.4$^e$ &8 
$\pm$ 3\\ 
Sex B$^\dagger$  & 7.84 $\pm$ 0.3 & 4.5 $\pm$ 0.6 & 30. & 3.49 & 34.8 &18\\ 
NGC6822$^\dagger$& 8.2  $\pm$ 0.2 & 13.4$\pm$ 1.8 & 164 & 2.02 &182.6$^e$&8\\ 
IC10$^\dagger$   & 8.19 $\pm$ 0.15& 15.3$\pm$ 3.5 & 158 & 1.94 
&95.86$^e$&8$^\ast$ $\pm$ 2\\ 
PegDig$^\dagger$ & 7.93 $\pm$ 0.14& 5.4 $\pm$ 0.6 & 5.8 & 0.62 & 26.4 
&8.6$^\ast$ $\pm$ 1.4\\
\hline
NGC 55$^\dagger$ & 8.32 $\pm$ 9.15& 139 $\pm$ 22.4& 1560& 14.12&317.4$^e$&8\\
NGC3109$^\dagger$& 8.06 $\pm$ 0.2 & 69  $\pm$ 14  & 655 & 7.56 &23.91$^e$&10 
$\pm$ 2\\
\end{tabular}
\end{center}
\tablenotetext{a}{from Skillman \etal 1989 or Mateo 1998; error estimates from 
Mateo 1998}
\tablenotetext{b}{from Huchtmeier and Richter 1988; or Mateo 1998}
\tablenotetext{c}{from Mateo 1998 or RC3}
\tablenotetext{\dagger}{Local Group galaxy (most numbers come from Mateo 1998)}
\tablenotetext{\ast}{more reliable $\sigma$ measures} 
\end{table}

\begin{table}
\begin{center}

\title{Table 2: The Van Zee Sample}
\vskip 0.5cm
\begin{tabular}{lccrccl}
\hline
Galaxy & Z\tablenotemark{a} & $M_{HI}$\tablenotemark{b} &
$M_{tot}$\tablenotemark{b}&$\varpi_*$\tablenotemark{b} & 
$<\Sigma_B>_{25}$\tablenotemark{b} & $\sigma_{ism}$\tablenotemark{b}\\ 
& 12+log(O/H) & $\times 10^7 M_\odot$&
$\times 10^{7}M_\odot$ & kpc & L$\odot pc^{-2}$ & km s$^{-1}$\\
\hline
\hline
UGC 300	 &    7.8$\pm$0.1  & 93$^\ast$  &  -  &    -     &    -    &  - \\
UGC 521  &    7.9$\pm$0.1  & 95$^\ast$  &  407$^\ast$ &   -    & - &	 - \\
\hline
UGCA 20   &   7.6$\pm$0.1  & 25   &   506 &  8.66   &  4.18  & 	9.0\\
UGC 2684  &   7.6$\pm$0.1  & 15   &   157 &  4.51   &  7.97  & 	7.6\\
UGC 2984  &   8.3$\pm$0.2  & 692$^\ast$ &     - &    -    &  20.0  & -\\
UGC 3174  &   7.8$\pm$0.1  & 61   &    911  &  9.34 &    31.7 & 9.2\\
UGC 5716  &   8.1$\pm$0.1  & 140  &    1820 &  13.67&    20.0 & 9.5\\
UGC 7178  &         -      & 140  &    1720 &  12.71 &   13.9 & 8.0 \\
UGC 11820 &   8.0$\pm$0.2  & 440  &    3910 &  21.21 &   22.0 & 8.6 \\
UGC 191   &   8.12$^c\pm$0.03 & 320 &    3250 &   15.97 &   55.2& 11.2 \\
UGC 634   &   8.18$^c\pm$0.03 & 470 &    6770 &  19.90  &  31.7 & 11.4\\
UGC 891   &   8.20$^c\pm$0.10 & 100 &    1080 &  11.48  &  24.1 & 8.8\\
UGC 5764  &   7.92$^c\pm$0.03 & 34  &    303  &  6.60   &  45.9 & 9.7\\
\hline
UGC 5829 &  8.28$^c\pm$0.10 & 182$^\ast$ & 2188$^\ast$ & - &  - &  -\\
UGC 8024   &  7.67$^c\pm$0.06 & 12$^\ast$  & 220$^\diamond$ &6$^\diamond$& - & 
 -\\
UGC 9128$^\dagger$& 7.75$^c\pm$0.05 & 3.5$^\ast$ &10$^\diamond$  
&1.76$^\diamond$&-&-\\
UGC 3672   &  8.0$\pm$  0.1  & 129$^\ast$ &    355$^\ast$ &   -     &    -  &  
-\\
UGCA 357   &  8.05$\pm$0.05  & 191$^\ast$ &    501$^\ast$ &   -     &    -  &  
-\\
Haro 43    &  8.20$\pm$0.1 &  229$^\ast$  &    295$^\ast$ &   -     &    -  &  
-\\
\hline
\end{tabular}
\end{center}
\tablenotetext{a}{from van Zee, Haynes \& Salzer (1997a) }
\tablenotetext{b}{from van Zee, Haynes, Salzer \& Broeils (1997)}
\tablenotetext{c}{from van Zee, Haynes \& Salzer (1997b)}
\tablenotetext{\ast}{from van Zee, Maddalena, Haynes, 
Hogg \& Roberts (1997)}
\tablenotetext{\diamond}{from Huchtmeier \& Richter 1988}
\tablenotetext{\dagger}{DDO187}
\end{table}

\small           

\begin{references}

Abbott, B. C., Bieging, J. H. \& Churchwell, E. 1981, ApJ, 250, 645

Arnett, D. W. 1978, ApJ, 219, 1008

Arnett, D. W. 1990, in Chemical and Dynamical Evolution of Galaxies,
eds. F.  Ferrini \etal (Pisa: ETS), 409

Babul, A. \& Rees, M. 1992, MNRAS, 253, 31

Babul, A. \& Ferguson, H. C. 1996, ApJ, 458, 100

Barkana, R. \& Loeb, A. 1999, preprint (astro-ph/9901114)

Binggeli, B. 1994, ESO/OHP Workshop on Dwarf Galaxies, eds. G. Meylan,
P. Prugniel (ESO: Garching), 123

Binney, J.  \& de Vaucouleurs, G. 1981, MNRAS, 194, 679

Binney, J. \& Tremaine, S. 1987, Galactic Dynamics (Princeton:
UniPress)

Bomans, D. J., Chu, Y.-H., \& Hopp, U. 1997, AJ, 113, 1678

Bowen, D. V., Tolstoy, E., Ferrara, A., Blades, J. C. \& Brinks, E.
1997, ApJ, 478, 530

Broeils, A. H. \& van Woerden, H. 1994, A\&AS, 107, 129

Burkert, A. 1995 ApJ, 447, L25

Burkert, A., Theis, C. \& Hensler, G. 1993  Dwarf Galaxies, ESO/OHP 
Workshop, eds.  Meylan \& Prugniel, 545 

Carignan, C. 1998, in Dwarf Galaxies and Cosmology, eds. Thuan \etal,
in press

Carignan, C., Beaulieu, S., C\^{o}t\'{e}, S., Demers, S. \& Mateo, M.
1998, AJ, 116, 1690

Casertano, S. P. R. \& Shostak, G. S. 1980, A\&A, 81, 371

Ciardi, B.  \& Ferrara, A. 1997, ApJL, 483, 5 

Ciardi, B., Ferrara, A. \& Abel, T. 1999, ApJ, in press 

Ciardi, B., Ferrara, A., Governato, F. \& Jenkins, A. 1999, in preparation 

Cole A. A., Tolstoy E., Gallagher J. S., Hoessel J. G., Mould J. R.,
Holtzmann J. A., Saha A. \& WFPC2 IDT 1999, AJ, in press

Couchman, H. M. P. \& Rees, M. J. 1986, MNRAS, 221, 53

Chyzy, K., Urbanik, M., Kohle, S., Klein, U., Beck, R. \& Berkhuijsen, E. M. 
1997, in
"Dwarf Galaxies: Probes for Galaxy Formation and Evolution", IAU JD

Clayton, D. D. \& Pantelaki, I. 1993, Phys. Reports, 227, 293

Davies, J.  \& Phillips, S. 1988, MNRAS, 233, 553

Dekel, A. \& Silk, J. 1986, ApJ, 303, 39

Della Ceca, R., Griffiths, R. E., Heckman, T. M. \& Mackenty,
J.W. 1996, ApJ, 469, 662

De Young, D. S. \& Gallagher, J. S. III 1990, ApJ, 356L, 15

De Young, D. S. \& Heckman, T. 1994, ApJ, 431, 598

Efsthatiou, G. 1992, MNRAS, 256, 43

Einasto, J., Saar, E., Kaasik, A. \& Chernin, A. D. 1974, Nature, 252,
111

Ferguson, H. C. \& Binggeli, B. 1994, A\&AR, 6, 67

Ferrara, A. 1998, ApJ, 499, L17 


Gallagher, J.S., Tolstoy, E., Saha, A., Hoessel, J.,
Dohm-Palmer, R.C., Skillman, E.D. \& Mateo, M.  1998, AJ, 115, 1869.

Haiman, Z., Rees, M. J., \& Loeb, A. 1997, ApJ, 484, 985

Han, M., Hoessel, J. G., Gallagher, J. S. III, Holtsman, J \& Stetson,
P. B. \& WFPC2/IDT 1997, AJ, 113, 1001

Heckman, T. H., Dahlem, M., Lehnert, M. D., Fabbiano, G., Gilmore, D.
\& Waller, W. H. 1995, ApJ, 448, 98

Hirashita, H., Takeuchi, T. T. \& Tamura, N. 1998, ApJL, 504, 83 

Huchtmeier, W. K. \& Richter, G. 1988, A\&A, 203, 237

Hunter, D. A., Gallagher, J. S. \& Rautenkranz, D. 1982, ApJS, 49, 53

Hunter, D. A., Wilcots, E. M., van Woerden, H., Gallagher, J. S.,
Kohle, S. 1998, ApJL, 495, 47


Klein, U., Giovanardi, C., Altschuler, D. R. \&  Wunderlich, E. 1992, A\&A, 
255, 49 

Klein, U., Hummel, E., Bomans, D. J. \& Hopp, U. 1996, A\&A, 313, 396

Klypin, A. A., Kravtsov, A., Valenzuela, O. \& Prada, F. 1999, ApJ, submitted
(astro-ph/9901240)

Knapp, G. R., Kerr, F. J. \& Bowers, P. F, 1978, AJ, 83, 360

Koeppen, J., Theis, C. \& Hensler, G. 1995, A\&A, 296, 99

Koo, B. C. \& McKee 1992, ApJ, 388, 103  

Kompaneets, A. S. 1960, Sov. Phys. Dokl., 5, 46

Larson, R. B.  1974, MNRAS, 166, 585

Lacey, C. \& Silk, J. 1991, ApJ, 381, 14

Lequeux, J, Rayo, J. F., Serrano, A., Peimbert, M. \& Torres-
Peimbert, S.  1979, A\&A, 80, 155

Lin, D. N. C. \& Faber, S. M. 1983, ApJL, 266, 21

Lin, D. N. C. \& Murray, S. D. 1992, ApJ, 394, 523

Lisenfeld, U. \& Ferrara, A. 1998, ApJ, 496, 145

Lo, K. Y., Sargent, W. L. W., Young, K. 1993, AJ, 106, 507

Mac~Low, M.-M., McCray, R. \& Norman, M. L. 1989, ApJ, 337, 141

Mac~Low, M.-M., \& Ferrara, A. 1999, ApJ, in press (astro-ph/9801237) (MF)

Maeder A., 1992, A\&A, 264, 105

Marconi, G., Matteucci, F. \& Tosi, M. 1994, MNRAS, 270, 35

Mateo, M. 1988, ApJ, 331, 261

Mateo, M. 1993, in Dwarf Galaxies, ESO/OHP Workshop, eds. 
Meylan \& Prugniel, 309 

Mateo, M. 1998, ARA\&A, 36, 435

Matteucci, F. \& Chiosi, C. 1983, A\&A, 123, 121

Matteucci, F. \& Tosi, M. 1985, MNRAS, 217, 391



Meurer, G. R., Freeman, K. C., Dopita, M. A., Cacciari, C. 1992, AJ, 103, 60 

Meurer, G. R. 1999, in Dwarf Galaxies and Cosmology, XIII Moriond Meeting, 
eds. T. X.
Thuan etal., in press (astro-ph/9806304)

Miniati, F., Jones, T. W., Ferrara, A. \& Ryu, D. 1997, ApJ, 491, 216

Miniati, F. Ryu, D., Ferrara, A. \& Jones, T. W. 1999, ApJ, 510, 726 

Minniti, D. \& Zijlstra, A. 1996, ApJL, 467, 13

Miralda-Escude, J. \& Rees, M. 1997, ApJ, 497, 21

Nagashima, M., Gouda, N. \& Sugiura, N. 1998, MNRAS 301, 849

Norman, C. A., \& Ferrara, A. 1996, ApJ, 467, 280

Norman, C. A., \& Spaans, M. 1997, ApJ, 480, 145

Padoan, P., Nordlund, A., Jones, B. J. T. 1997, MNRAS, 288, 43

Pagel, B. E. J., Simonson, E. A., Terlevich, R. J. \& Edmunds, M. G. 1992,
MNRAS, 255, 325

Papaderos, P., Fricke, K. J., Thusn, T. X. \& Loose, H. -H. 1994,
A\&A, 291, L13

Patterson, R. \& Thuan, T. 1996, ApJSS, 107, 103

Persic, M., Salucci, P. \& Stel, F. 1996, MNRAS, 281, 27P


Pryor, C. 1996, The Formation of the Galactic Halo: Inside and Out,
ASP Conf.  Ser., Vol. 92, eds. H. Morrison and A. Sarajedini, (San
Francisco: PASP), 424

Puche, D., Westphal, D., Brinks, E. \& Roy, J.-R. 1992, AJ, 103, 1841

Puche, D. \& Westphal, D. 1993, in Dwarf Galaxies, ESO/OHP Workshop, eds.
Meylan \& Prugniel, 273
 
Richer, M. G., McCall, M. L. \& Stasinka, G. 1999, A\&A in press 
(astro-ph/9809302)

Ricotti, M., Ferrara, A. \& Miniati, F. 1997, ApJ, 485, 254

Salpeter, E. E. \& Hoffman, G. L. 1996, ApJ, 465, 595

Scalo, J. M. 1986, Fund. Cosm. Phys., 11, 1

Silk, J. 1997, ApJ, 481, 703

Skillman E. D., Melnick, J., Terlevich, R. \& Moles, M. 1988, A\&A,
196, 31

Skillman, E. D., Kennicutt, R.C. \& Hodge, P. W. 1989, ApJ, 347, 875

Smecker-Hane, T. A., Stetson, P. B., Hesser, J. E. \& Lehnert,
M. D. 1994, AJ, 108, 507

Spaans, M. \& Norman, C. A. 1997, ApJ, 483, 87

Spitzer, L. 1978, Physical Processes in the Interstellar Medium, (New
York: Wiley)

Staveley-Smith, L., Davies, R. D. \& Kinman, T. D. 1992, MNRAS, 258,
334


Suntzeff, N. B., Mateo, M., Terndrup, D. M., Olszewski, E. W.,
Geisler, D.
\& Weller, W. 1993, ApJ, 418, 208

Taylor, C. L. 1997, ApJ, 480, 524 

Tegmark, M., Silk, J., Rees, M.J., Blanchard, A., Abel, T. \& Palla, F.
1997, ApJ, 474, 1

Thielemann, F. K., Nomoto, K. \& Hashimoto, M. 1992, Structure and
Evolution of Neutron Stars, (Knudsen), 298

Thoul, A. \& Weinberg, D. H. 1996, ApJ, 465, 608

Tinsley, B. M. 1980, Fund. Cosm. Phys., 5, 287

Tolstoy, E. 1996, ApJ, 462, 684

Tolstoy, E. \& Saha, A. 1996, ApJ, 462, 672

Tolstoy, E., Gallagher, J.S., Saha, A., Hoessel, J.,
Skillman, E.D., Dohm-Palmer, R.C., Mateo, M. \& Hurley-Keller D.
1998, AJ, 116, 1244

Tolstoy, E. 1998a, in Dwarf Galaxies and Cosmology, eds. Thuan \etal,
in press (astro-ph/9807154)

Tolstoy, E. 1998b, in The Stellar Content of Local Group Galaxies, 
eds. Whitelock \& Cannon, in press (astro-ph/9901245)

Vader, J. P. 1986, ApJ, 305, 669
 
Vader, J. P. 1987, ApJ, 317, 128


van Zee, L., Haynes, M. P. \& Salzer, J. J. 1997a, AJ, 114, 2479

van Zee, L., Haynes, M. P. \& Salzer, J. J. 1997b, AJ, 114, 2497

van Zee, L., Haynes, M. P., Salzer, J. J., \& Broeils, A. 1997, AJ, 113, 1618

van Zee, L., Maddalena R. J., Haynes, M. P., Hogg, D. E. \& 
Roberts, M. S. 1997, AJ, 113, 1638

van Zee, L., Skillman, E.D. \& Salzer, J. J. 1998, AJ, 116, 1186

Wang, B. \& Silk, J.  1993, ApJ, 406, 580


Wilcots, E.M. \& Miller B.W. 1998, AJ, 116, 2363

Wyse, R. F. G.  \& Silk, J. 1985, ApJL, 269, 1

Woosley, S. E. \& Weaver, T. A. 1986, ARA\&A, 24, 205

Young, L. M.  \& Lo, K. Y. 1996, ApJ, 462, 203

Young, L. M.  \& Lo, K. Y. 1997a, ApJ, 476, 127




\end{references}
\end{document}